\documentclass[fleqn,usenatbib]{mnras}

\usepackage{newtxtext,newtxmath}
\usepackage[T1]{fontenc}

\DeclareRobustCommand{\VAN}[3]{#2}
\let\VANthebibliography\thebibliography
\def\thebibliography{\DeclareRobustCommand{\VAN}[3]{##3}\VANthebibliography}

\usepackage{graphicx}
\usepackage{amsmath}

\title[Rotating BHs in dark matter halos]
{Adaptive ray tracing, image diagnostics, and photon ring signatures of rotating dark-matter-dressed black holes}

\author[M. Fathi]{
Mohsen Fathi,$^{1}$\thanks{E-mail: mohsen.fathi@ucentral.cl}
\\
$^{1}$Centro de Investigaci\'{o}n en Ciencias del Espacio y F\'{i}sica Te\'{o}rica (CICEF),
Universidad Central de Chile, La Serena 1710164, Chile
}

\date{Accepted XXX. Received YYY; in original form ZZZ}

\pubyear{\the\year{}}

\begin{document}
\label{firstpage}
\pagerange{\pageref{firstpage}--\pageref{lastpage}}
\maketitle

%%%%%%%%%%%%%%%%%%%%%%%%%%%%%%%%%%%%%%%%%%%%%%%%%%
% Abstract
%%%%%%%%%%%%%%%%%%%%%%%%%%%%%%%%%%%%%%%%%%%%%%%%%%

\begin{abstract}
We study the optical appearance of rotating black holes embedded in dark matter environments using a phenomenological ray  tracing framework. Rather than focusing on a single geometry, we compare two effective rotating backgrounds obtained from static dark matter sourced seed metrics: a regular Einasto-type black hole and a cored-NFW black hole. Kerr is used as the reference spacetime. We construct observer-screen images by numerical backward ray tracing and analyse the horizon structure, shadow boundary, lensing bands, transfer maps, and synthetic intensity distributions produced by a common semi-analytic accretion prescription. We also introduce simple image-level diagnostics, an angular-scale confrontation with M87* and Sgr A*, and simplified visibility-amplitude diagnostics. These additions are not intended as an EHT fit, but as a controlled way to identify which observables are most affected by the dark matter dressing. For the representative parameters considered here, the Einasto-supported geometry remains very close to Kerr, while the cored-NFW case produces a stronger redistribution of the image, with larger centroid displacement, stronger brightness asymmetry, an outward shift of the characteristic bright-ring scale, and a visible change in the normalized visibility amplitude. The results indicate that rotating dark-matter-dressed backgrounds can produce systematic image-domain and Fourier-domain deviations that are partially degenerate with spin, inclination, and emission modelling. The framework is lightweight and extensible, and provides a first step toward future GRRT and GRMHD studies of rotating black holes in dark matter environments.
\end{abstract}

%%%%%%%%%%%%%%%%%%%%%%%%%%%%%%%%%%%%%%%%%%%%%%%%%%
% Keywords
%%%%%%%%%%%%%%%%%%%%%%%%%%%%%%%%%%%%%%%%%%%%%%%%%%


%%%%%%%%%%%%%%%%%%%%%%%%%%%%%%%%%%%%%%%%%%%%%%%%%%
%%%%%%%%%%%%%%%%% BODY OF PAPER %%%%%%%%%%%%%%%%%%
%%%%%%%%%%%%%%%%%%%%%%%%%%%%%%%%%%%%%%%%%%%%%%%%%%

%%%%%%%%%%%%%%%%%%%%%%%%%%%%%%%%%%%%% Sect.1
\section{Introduction}

The observational study of black holes has entered a new stage after the horizon-scale images of M87* and Sgr A* reported by the Event Horizon Telescope (EHT) Collaboration \citep{EHT2019I,EHT2022I}. These observations have shown that the optical appearance of a black hole is shaped not only by the presence of an event horizon, but also by light bending, photon region structure, and the properties of the surrounding emitting plasma. In this context, the shadow, the bright emission ring, and possible higher-order photon ring contributions provide complementary probes of the strong-field geometry \citep{Falcke2000,Johannsen2010,Bambi2013,GrallaHolzWald2019,Johnson2020}. Although the Kerr metric remains the standard background for interpreting compact-object images in General Relativity, astrophysical black holes are not isolated. They are embedded in environments where baryonic matter, plasma, magnetic fields, and dark matter may affect the observed image.

Dark matter plays a central role in the formation and evolution of galaxies and clusters. Several density profiles have been proposed to describe galactic dark matter distributions, including the Navarro--Frenk--White (NFW) profile, the Einasto profile, cored profiles, and other phenomenological halo models \citep{Navarro1996,Navarro1997,Einasto1965,Cardone2005,DiCintio2014}. In ordinary astrophysical situations, the direct influence of a galactic halo on the near-horizon region is expected to be small. Nevertheless, dark-matter-dressed black hole solutions provide a useful phenomenological setting for asking how environmental matter distributions may modify geodesic motion, lensing observables, and accretion images. They also make it possible to explore degeneracies between the parameters of the central compact object and those describing the surrounding dark sector.

From the theoretical side, dark matter environments have been used to construct black hole spacetimes beyond the Schwarzschild or Kerr backgrounds. Some works treat the halo as an external matter distribution, while others incorporate the matter profile directly into the spacetime geometry. In this direction, several profiles, such as Hernquist, Dehnen, Einasto, and cored halo models, have been used to analyse shadows, photon spheres, quasinormal modes, thermodynamics, and EHT-related constraints \citep{Xu2018,Hou2018,Hou2018PFDM,Cardoso2022,Feng2026,Senjaya2025Dehnen,FathiCruz2023,Fathi2025Universe,FathiAhmed2026}. A relevant recent development is the construction of exact black hole geometries in which the dark matter distribution acts as an effective source of the spacetime itself. In particular, regular black hole solutions sourced by galactic dark matter profiles, including Einasto-type distributions, have been proposed \citep{KonoplyaZhidenko2026}. Exact black hole solutions immersed in cored-NFW halos have also been analysed in connection with energy conditions, null geodesics, optical properties, perturbations, and thermodynamics \citep{Senjaya2026}. These models provide a natural starting point for asking whether different dark matter profiles can leave distinguishable signatures in the optical appearance of rotating black holes.

Most existing studies of dark-matter-dressed black holes have focused on static and spherically symmetric configurations, or on the shadow boundary after applying a Kerr-like rotating extension. Such analyses are important, but the shadow diameter alone is not enough to disentangle dark matter effects from changes in spin, mass, inclination, or emission model. A more informative route is to go beyond the geometrical shadow contour and study the full ray-traced image, including the intensity distribution, lensing bands, higher-order images, and photon-ring-related structures.

Photon rings are important because they are associated with photons that execute multiple half-orbits around the black hole before reaching the observer. Their properties are controlled by near-critical geodesics and are therefore sensitive to the strong-field spacetime structure \citep{GrallaHolzWald2019,Johnson2020,GrallaLupsasca2020}. Adaptive ray  tracing techniques provide an efficient way to resolve these narrow features. In particular, Adaptive Analytical ray tracing was introduced to compute high-resolution Kerr photon ring images and visibility signatures by exploiting separability \citep{CardenasAvendano2023}. For more general rotating backgrounds, separability need not be available. The same adaptive idea can nevertheless be implemented numerically by refining the observer screen near the shadow boundary, the critical curve, and the higher-order image regions.

The aim of this work is to develop a lightweight numerical framework for the optical appearance of rotating black holes dressed by dark matter profiles. We consider two effective rotating geometries generated from physically motivated static seeds: a regular Einasto-type black hole and a cored-NFW black hole. Kerr is used as the reference spacetime. Since Newman--Janis-type procedures do not uniquely determine the rotating matter sector, the rotating metrics are treated here as phenomenological Kerr-like backgrounds rather than unique exact rotating solutions of the original matter distribution. This keeps the interpretation conservative while still allowing us to probe how different dark matter profiles affect ray-traced images.

The main purpose is not to claim a dark matter detection from black hole images. Instead, we ask how the dressing changes the geometrical and image-domain observables that enter horizon-scale imaging. We therefore complement the shadow and transfer-map analysis with synthetic images, image-level morphology diagnostics, an angular-scale confrontation with the reference scales of M87* and Sgr A*, and simplified visibility-amplitude diagnostics. These diagnostics are meant as a bridge between the spacetime geometry and the quantities that would eventually enter a more complete general relativistic radiative-transfer (GRRT) or general relativistic magnetohydrodynamic (GRMHD) modelling pipeline.

Our analysis is based on numerical backward ray tracing. We integrate null geodesics from an observer screen toward the compact object and classify the rays according to capture, escape, and intersection with the emitting region. We then use the image-plane structure to identify the shadow boundary, lensing bands, and transfer maps with moderate computational cost. The emission is described by a semi-analytic optically thin disk-like prescription. This model is not meant to replace GRRT or GRMHD calculations; it is used to apply the same controlled emissivity to Kerr, Einasto-supported, and cored-NFW geometries and thereby isolate the role of the spacetime.

The main questions addressed in this paper are the following. First, we study how the horizon, ergoregion, and shadow boundary change when the rotating black hole is dressed by an Einasto-type or cored-NFW profile. Second, we investigate whether the dark matter parameters mainly affect the primary shadow size, or whether their impact is more visible in lensing bands, transfer maps, higher-order images, and photon-ring-related structures. Third, we examine the degeneracy between spin, inclination, emission prescription, and dark matter parameters. Finally, we introduce compact image-domain and Fourier-domain diagnostics, and use them for a limited angular-scale confrontation with M87* and Sgr A*, without performing an EHT fit.

The paper is organized as follows. In Section~\ref{sec:geometry}, we introduce the static dark-matter-dressed seed geometries and construct their effective rotating extensions. In Section~\ref{sec:geodesics}, we present the Hamiltonian formulation of null geodesics and the observer-screen setup. In Section~\ref{sec:adaptive_ray_tracing}, we describe the adaptive ray  tracing procedure, lensing bands, and transfer maps. In Section~\ref{sec:images}, we introduce the semi-analytic emission model and compute the synthetic images. In Section~\ref{sec:phenomenology}, we discuss the phenomenological implications, morphology diagnostics, angular-scale confrontation, visibility diagnostics, and limitations. Finally, in Section~\ref{sec:conclusions}, we summarize the main results and outline future extensions.

%%%%%%%%%%%%%%%%%%%%%%%%%%%%%%%%%%%%% Sect.2
\section{Rotating dark-matter-dressed black hole geometries}
\label{sec:geometry}

In this section, we introduce the class of dark-matter-dressed black hole geometries that will be used as the background for the ray  tracing analysis. The construction is organized in two steps. First, we specify two static and spherically symmetric seed metrics associated with physically motivated dark matter profiles. Second, we promote these seed geometries to effective rotating backgrounds through a Kerr-like extension. Throughout this work, we use geometrized units, $G=c=1$.

To avoid dimensional ambiguities in the numerical analysis, we work with the dimensionless variables
\begin{equation}
T=\frac{t}{M},\qquad
R=\frac{r}{M},\qquad
\chi=\frac{a}{M}=\frac{J}{M^2},
\label{eq:dimensionless_basic_variables}
\end{equation}
where $M$ is the asymptotic gravitational mass and $\chi$ is the dimensionless spin parameter. We also define the dimensionless line element
\begin{equation}
d\sigma^2=\frac{ds^2}{M^2}.
\label{eq:dimensionless_line_element}
\end{equation}
In this notation, $R$ denotes the radial coordinate. Later, in the geodesic section, we will use a different symbol, $q^\mu$, for the spacetime coordinates in order to avoid confusion with the dimensionless radius.

The first seed is the regular Einasto-supported black hole constructed by Konoplya and Zhidenko, in which the dark matter distribution is treated as an effective anisotropic source and the condition $P_r=-\rho$ leads to a Schwarzschild-like gauge with $g_{tt}g_{rr}=-1$ \citep{KonoplyaZhidenko2026}. The second seed is the cored-NFW black hole solution proposed in \citet{Senjaya2026}, where the metric function is obtained from the cored-NFW halo distribution and the corresponding tangential velocity profile. We stress that these two geometries are not superposed into a single matter distribution. Instead, they are treated as two independent members of the same phenomenological class of dark-matter-dressed black holes. This comparative construction allows us to identify which optical features are profile dependent, while avoiding the introduction of an artificial combined matter source.

%%%%%%%%%%
\subsection{Static seed geometries}
\label{subsec:static-seeds}

In dimensionless variables, the static and spherically symmetric seed geometry is written as
\begin{equation}
d\sigma^2
=
-H_i(R)\,dT^2
+\frac{dR^2}{H_i(R)}
+R^2\left(d\theta^2+\sin^2\theta\,d\phi^2\right),
\label{eq:static_seed_dimensionless}
\end{equation}
where the index $i$ labels the dark matter model. In this work, we use the model-indexed lapse
\begin{equation}
H_i(R)=
\begin{cases}
H_{\rm E}(R), & i={\rm E},\\[2mm]
H_{\rm cNFW}(R), & i={\rm cNFW}.
\end{cases}
\label{eq:model_indexed_lapse_dimensionless}
\end{equation}
This notation should be understood as a compact way of treating two separate seed metrics within the same ray  tracing framework. It does not represent a single spacetime containing both dark matter profiles simultaneously.

For the Einasto-supported case, the reliability of the static seed follows from its direct construction through the Einstein field equations. In the original dimensional form, the source is modelled as an anisotropic fluid,
\begin{equation}
T^{\mu}_{\ \nu}
=
\mathrm{diag}[-\rho(r),P_r(r),P_t(r),P_t(r)],
\label{eq:anisotropic_tensor}
\end{equation}
and the condition $P_r=-\rho$ leads to the Schwarzschild-like gauge used in Eq.~\eqref{eq:static_seed_dimensionless} \citep{KonoplyaZhidenko2026}. In dimensionless form, if we define
\begin{equation}
\mu(R)=\frac{m(r)}{M},
\label{eq:dimensionless_mass_function_general}
\end{equation}
the Einstein equation for the mass function becomes
\begin{equation}
\frac{d\mu}{dR}
=
4\pi M^2 R^2 \rho(MR),
\label{eq:dimensionless_mass_density_relation}
\end{equation}
and the lapse is fixed by
\begin{equation}
H(R)=1-\frac{2\mu(R)}{R}.
\label{eq:dimensionless_lapse_mass_relation}
\end{equation}
Thus, once the density profile is specified, the static metric is determined by the Einstein equations rather than introduced as an arbitrary deformation of Schwarzschild.

For the analytic Einasto-type case considered here, the dimensionless lapse is \citep{KonoplyaZhidenko2026}
\begin{equation}
H_{\rm E}(R)
=
1-\frac{2}{R}
+
\left(
\frac{2}{R}
+\frac{2}{\lambda_{\rm E}}
+\frac{R}{\lambda_{\rm E}^{2}}
\right)
\exp\left(-\frac{R}{\lambda_{\rm E}}\right),
\label{eq:h_einasto_dimensionless}
\end{equation}
where
\begin{equation}
\lambda_{\rm E}=\frac{\ell_{\rm E}}{M}
\label{eq:lambda_e_def}
\end{equation}
is the dimensionless Einasto scale. In the static configuration, this geometry is regular at the origin because the mass function vanishes sufficiently fast as $R\rightarrow 0$. At large radius, $H_{\rm E}(R)\rightarrow 1-2/R$, and the Schwarzschild behaviour is recovered asymptotically.

The second seed is the cored-NFW black hole geometry of \citet{Senjaya2026}. In terms of the dimensionless radius $R$, the halo density profile can be written as
\begin{equation}
\rho_{\rm cNFW}(R)
=
\rho_0
\left(
1+\frac{R}{\lambda_{\rm c}}
\right)^{-3},
\label{eq:cored_nfw_density_dimensionless_radius}
\end{equation}
where
\begin{equation}
\lambda_{\rm c}=\frac{r_0}{M}
\label{eq:lambda_c_def}
\end{equation}
is the dimensionless core radius. The corresponding dimensionless lapse is
\begin{equation}
H_{\rm cNFW}(R)
=
\exp\left\{
\eta \lambda_{\rm c}
\left[
\frac{1}{R+\lambda_{\rm c}}
-\frac{2}{R}\ln\left(1+\frac{R}{\lambda_{\rm c}}\right)
\right]
\right\}
-\frac{2}{R},
\label{eq:h_cnfw_dimensionless}
\end{equation}
with
\begin{equation}
\eta=4\pi\rho_0 r_0^2 .
\label{eq:eta_def}
\end{equation}
In the construction of \citet{Senjaya2026}, the cored-NFW density profile fixes the enclosed dark matter mass and the associated tangential velocity, from which the metric function is obtained. The energy conditions of the resulting static spacetime were also analysed in that work. Here, this solution is used as an independent static seed in order to compare its optical signatures with those of the regular Einasto-supported black hole.

For both seed metrics, it is useful to define the dimensionless effective radial mass function
\begin{equation}
\mu_i(R)
=
\frac{R}{2}\left[1-H_i(R)\right].
\label{eq:effective_mass_function_dimensionless}
\end{equation}
In the Schwarzschild limit, $\mu_i(R)=1$. For the dark-matter-dressed cases, $\mu_i(R)$ becomes radius dependent and encodes the deviation from the vacuum geometry. This quantity will be used below to write the rotating extension in a compact form.

\begin{figure}
\centering
\includegraphics[width=\columnwidth]{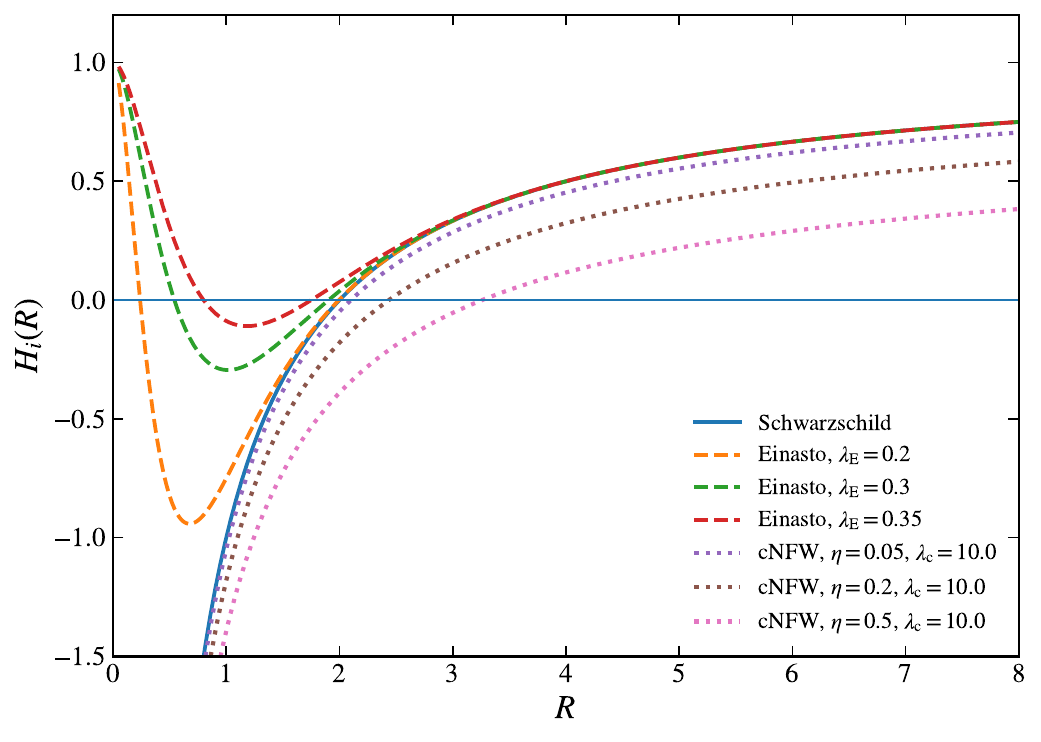}
\caption{
Static lapse functions $H_i(R)$ for the regular Einasto-supported black hole and the cored-NFW black hole geometry, compared with the Schwarzschild lapse $H_{\rm Schw}(R)=1-2/R$. The parameter values are chosen as representative examples to show the qualitative effect of the dark matter dressing. The zeros of the lapse functions indicate the corresponding static horizon locations.
}
\label{fig:static_lapses}
\end{figure}

The static lapse functions are shown in Fig.~\ref{fig:static_lapses} for representative parameter values. The purpose of this figure is not to impose observational bounds, but to illustrate how the Einasto-supported and cored-NFW seed geometries depart from the Schwarzschild reference. The zeros of $H_i(R)$ identify the static horizon positions, while the asymptotic approach to $1-2/R$ shows that the Schwarzschild behaviour is recovered at large radius.

%%%%%%%%%%
\subsection{Effective rotating extension}
\label{subsec:rotating-extension}

We now construct an effective rotating extension of the static seed metrics. The procedure is inspired by the Newman--Janis algorithm, originally introduced as a complex-coordinate prescription that generates the Kerr metric from the Schwarzschild spacetime and the Kerr--Newman metric from the Reissner--Nordstr\"om solution \citep{NewmanJanis1965,NewmanCouch1965}. Although the algorithm is remarkably successful in these standard cases, its application to generic matter sources is not unique. In particular, the complexification step can introduce ambiguities, and the resulting rotating metric is not guaranteed to solve the same field equations or to be sourced by the same matter sector as the original static seed \citep{DrakeSzekeres2000,LimaJunior2020}. For this reason, we use a modified Newman--Janis-type construction in the spirit of the non-complexification approaches developed for regular and effective black hole geometries \citep{BambiModesto2013,AzregAinou2014,Toshmatov2014}. The resulting spacetime is therefore interpreted as a phenomenological Kerr-like background rather than as the unique exact rotating solution of the original dark matter fluid.

To outline the construction in the dimensionless variables introduced above, we first introduce an Eddington--Finkelstein-type null coordinate $U$ through
\begin{equation}
dT=dU+\frac{dR}{H_i(R)} .
\label{eq:ef_coordinate_dimensionless}
\end{equation}
Then Eq.~\eqref{eq:static_seed_dimensionless} takes the form
\begin{equation}
d\sigma^2
=
-H_i(R)\,dU^2
-2\,dU\,dR
+R^2d\Omega^2 ,
\label{eq:static_seed_null_dimensionless}
\end{equation}
where $d\Omega^2=d\theta^2+\sin^2\theta\,d\phi^2$. The inverse metric can be represented by the null tetrad
\begin{align}
l^\mu &= \delta^\mu_R,
\label{eq:tetrad_l_dimensionless}
\\
n^\mu &= \delta^\mu_U-\frac{H_i(R)}{2}\delta^\mu_R,
\label{eq:tetrad_n_dimensionless}
\\
m^\mu &=
\frac{1}{\sqrt{2}\,R}
\left(
\delta^\mu_\theta
+\frac{i}{\sin\theta}\delta^\mu_\phi
\right),
\label{eq:tetrad_m_dimensionless}
\end{align}
together with its complex conjugate $\bar{m}^{\mu}$. With the signature convention adopted here, the contravariant metric is reconstructed as
\begin{equation}
g^{\mu\nu}
=
-l^\mu n^\nu-l^\nu n^\mu
+m^\mu \bar{m}^{\nu}
+m^\nu \bar{m}^{\mu}.
\label{eq:inverse_metric_tetrad_dimensionless}
\end{equation}
The Newman--Janis prescription introduces the complex transformation
\begin{equation}
U\rightarrow U-i\chi\cos\theta,
\qquad
R\rightarrow R+i\chi\cos\theta.
\label{eq:nj_complex_transformation_dimensionless}
\end{equation}
In the modified version adopted here, the dark matter functions are not arbitrarily complexified. Instead, the radial mass profile $\mu_i(R)$ is kept as a real function of $R$, while the angular dependence enters through the Kerr-like replacement
\begin{equation}
R^2\rightarrow \Sigma,
\qquad
\Sigma=R^2+\chi^2\cos^2\theta .
\label{eq:sigma_def_dimensionless}
\end{equation}
The Schwarzschild quantity $2R$ appearing in the dimensionless Kerr metric is then generalized to
\begin{equation}
\mathcal{F}_i(R)
=
2R\,\mu_i(R)
=
R^2\left[1-H_i(R)\right],
\label{eq:F_def_dimensionless}
\end{equation}
and the radial function becomes
\begin{equation}
\Delta_i(R)
=
R^2H_i(R)+\chi^2
=
R^2-2R\,\mu_i(R)+\chi^2 .
\label{eq:delta_def_dimensionless}
\end{equation}
After reconstructing the metric and performing a Boyer--Lindquist-type coordinate transformation to remove the $dR\,dT$ and $dR\,d\phi$ cross terms, one obtains
\begin{eqnarray}
d\sigma^2 &=&
-\left(1-\frac{\mathcal{F}_i(R)}{\Sigma}\right)dT^2
-\frac{2\chi\,\mathcal{F}_i(R)\sin^2\theta}{\Sigma}\,dT\,d\phi
+\frac{\Sigma}{\Delta_i(R)}\,dR^2
\nonumber\\
&&+\Sigma\,d\theta^2+
\sin^2\theta
\left[
R^2+\chi^2
+\frac{\chi^2\mathcal{F}_i(R)\sin^2\theta}{\Sigma}
\right]d\phi^2 .
\label{eq:rotating_metric_dimensionless}
\end{eqnarray}
This metric has the same algebraic structure as the Kerr geometry, but with the constant Kerr mass replaced by the radial mass function associated with the chosen dark-matter-dressed seed. If $H_i(R)=1-2/R$, then $\mathcal{F}_i(R)=2R$ and $\Delta_i(R)=R^2-2R+\chi^2$, so Eq.~\eqref{eq:rotating_metric_dimensionless} reduces to the Kerr metric in dimensionless Boyer--Lindquist coordinates. On the other hand, if $\chi\rightarrow 0$, then $\Sigma\rightarrow R^2$, $\Delta_i(R)\rightarrow R^2H_i(R)$, and Eq.~\eqref{eq:rotating_metric_dimensionless} reduces to the static seed metric \eqref{eq:static_seed_dimensionless}.

The nonvanishing dimensionless metric components are therefore
\begin{align}
g_{TT} &=
-\left(1-\frac{\mathcal{F}_i(R)}{\Sigma}\right),
\label{eq:gTT_rot_dimensionless}
\\
g_{T\phi} &=
-\frac{\chi\,\mathcal{F}_i(R)\sin^2\theta}{\Sigma},
\label{eq:gTphi_rot_dimensionless}
\\
g_{RR} &= \frac{\Sigma}{\Delta_i(R)},
\label{eq:gRR_rot_dimensionless}
\\
g_{\theta\theta} &= \Sigma,
\label{eq:gthetatheta_rot_dimensionless}
\\
g_{\phi\phi} &=
\sin^2\theta
\left[
R^2+\chi^2
+\frac{\chi^2\mathcal{F}_i(R)\sin^2\theta}{\Sigma}
\right].
\label{eq:gphiphi_rot_dimensionless}
\end{align}
These components will be used in the Hamiltonian formulation of the null geodesic equations. Notice that, for the dark-matter-dressed cases, $\Delta_i(R)$ is not generally a quadratic polynomial. Therefore, the horizon structure and the photon dynamics must be analysed numerically.

A formal effective matter source for the rotating geometry can be defined through
\begin{equation}
T^{\rm eff}_{\mu\nu}
=
\frac{1}{8\pi}G_{\mu\nu}.
\label{eq:effective_Tmunu}
\end{equation}
However, the detailed microphysical interpretation of this source is beyond the scope of the present work. Our aim is instead to use Eq.~\eqref{eq:rotating_metric_dimensionless} as a controlled phenomenological geometry for comparing the ray-traced optical signatures of different dark-matter-dressed black hole backgrounds.

\begin{figure}
\centering
\includegraphics[width=\columnwidth]{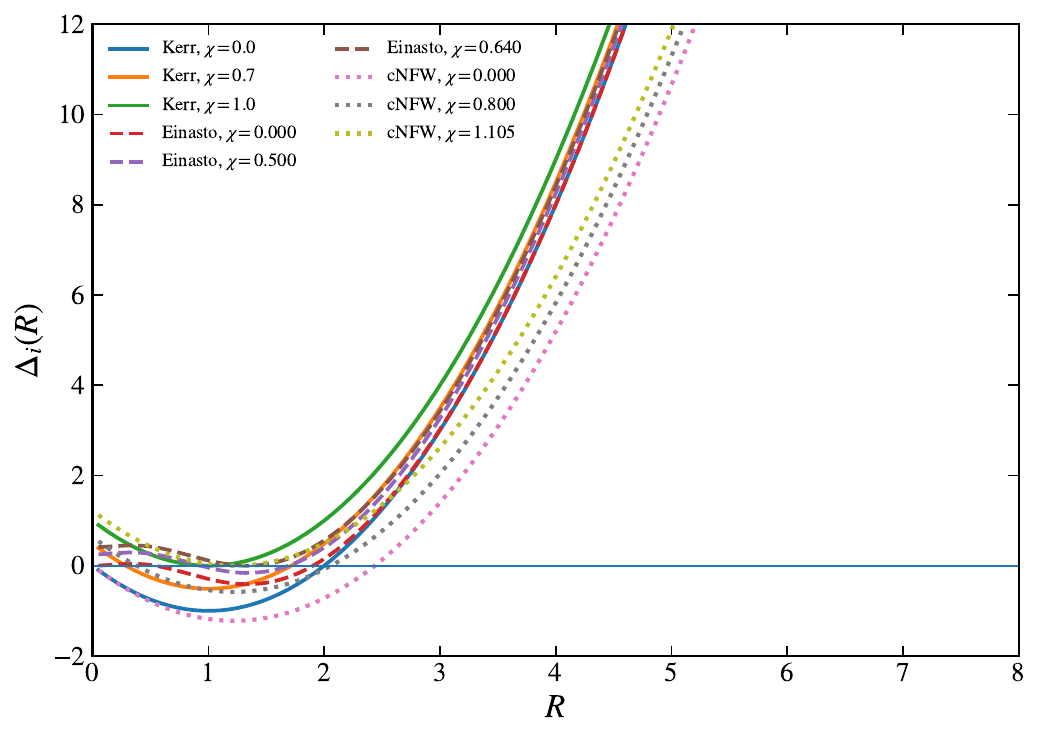}
\caption{
Rotating horizon function $\Delta_i(R)$ for the Kerr reference spacetime and for representative Einasto-supported and cored-NFW configurations. The zeros of $\Delta_i(R)$ determine the horizon radii. The extremal cases correspond to curves that touch the horizontal axis without crossing it. For the representative choices shown here, they occur at $\chi_{\rm e}=1$, $0.640$, and $1.105$ for the Kerr, Einasto-supported, and cored-NFW cases, respectively.
}
\label{fig:delta_function}
\end{figure}

The rotating radial function $\Delta_i(R)$ is displayed in Fig.~\ref{fig:delta_function}. Since the horizons are given by the roots of $\Delta_i(R)=0$, this diagnostic makes the subextremal and extremal cases transparent. In particular, a subextremal black hole corresponds to two positive roots, while in the extremal case the curve touches the horizontal axis.

%%%%%%%%%%
\subsection{Horizons and ergoregions}
\label{subsec:horizon_ergoregion}

The horizons of the rotating dark-matter-dressed geometries are determined by the coordinate singularities of $g_{RR}$, or equivalently by the roots of
\begin{equation}
\Delta_i(R)
=
R^2H_i(R)+\chi^2
=0.
\label{eq:horizon_condition_dimensionless}
\end{equation}
The outer event horizon, when it exists, is denoted by $R_+$ and corresponds to the largest positive root of Eq.~\eqref{eq:horizon_condition_dimensionless}. Depending on the spin and the dark matter parameters, this equation may admit two horizons, one degenerate horizon, or no horizon. The degenerate case is obtained from
\begin{equation}
\Delta_i(R_{\rm e})=0,
\qquad
\left.\frac{d\Delta_i}{dR}\right|_{R=R_{\rm e}}=0,
\label{eq:extremal_condition_dimensionless}
\end{equation}
where $R_{\rm e}$ denotes the extremal horizon radius. Since
\begin{equation}
\frac{d\Delta_i}{dR}
=
2R H_i(R)+R^2H_i'(R),
\label{eq:delta_derivative_dimensionless}
\end{equation}
the extremal radius satisfies
\begin{equation}
2H_i(R_{\rm e})+R_{\rm e}H_i'(R_{\rm e})=0,
\label{eq:extremal_radius_condition}
\end{equation}
and the corresponding extremal spin is given by
\begin{equation}
\chi_{\rm e}^{2}
=
-R_{\rm e}^{2}H_i(R_{\rm e}).
\label{eq:extremal_spin_condition}
\end{equation}
Here and in what follows, a prime denotes differentiation with respect to $R$. These relations determine the boundary between the black hole and horizonless regions in the parameter space.

\begin{figure}
\centering
\includegraphics[width=\columnwidth]{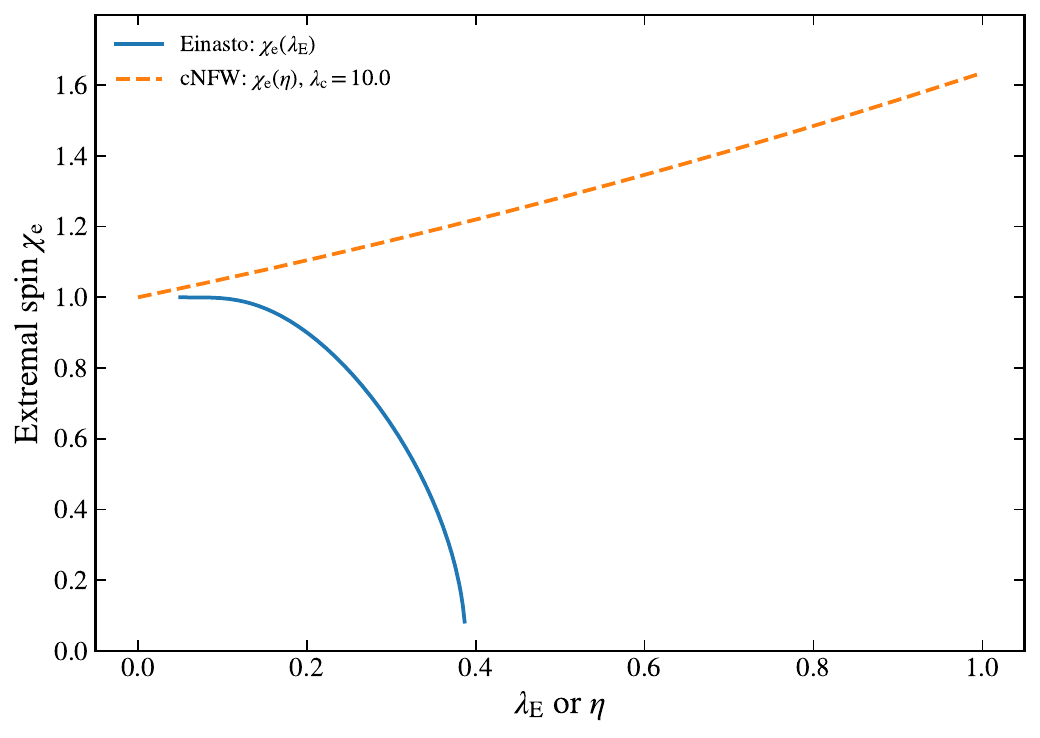}
\caption{
Extremal spin parameter $\chi_{\rm e}$ for the rotating Einasto-supported and cored-NFW geometries. The curves are obtained from $\Delta_i(R_{\rm e})=0$ and $\Delta_i'(R_{\rm e})=0$. They separate the black hole domain from the horizonless region in the corresponding parameter space.
}
\label{fig:extremal_spin}
\end{figure}

Figure~\ref{fig:extremal_spin} shows the extremal spin obtained from Eqs.~\eqref{eq:extremal_radius_condition} and \eqref{eq:extremal_spin_condition}. Configurations below the corresponding extremal curve have an outer horizon, whereas configurations above it are horizonless. This scan is used only to identify the domain in which the subsequent ray  tracing calculations describe black hole spacetimes.

The stationary-limit surface, or ergosurface, is determined by
\begin{equation}
g_{TT}=0.
\label{eq:ergosurface_condition_1_dimensionless}
\end{equation}
Using Eq.~\eqref{eq:gTT_rot_dimensionless}, this condition becomes
\begin{equation}
\Sigma-\mathcal{F}_i(R)=0,
\label{eq:ergosurface_condition_2_dimensionless}
\end{equation}
or equivalently
\begin{equation}
R^2H_i(R)+\chi^2\cos^2\theta=0.
\label{eq:ergosurface_condition_3_dimensionless}
\end{equation}
The outer stationary-limit surface is denoted by $R_{\rm s}(\theta)$ and is given by the largest positive solution of Eq.~\eqref{eq:ergosurface_condition_3_dimensionless} at fixed polar angle. At the poles, $\theta=0,\pi$, Eq.~\eqref{eq:ergosurface_condition_3_dimensionless} reduces to the horizon equation \eqref{eq:horizon_condition_dimensionless}, so the ergosurface touches the outer horizon as in Kerr. In the equatorial plane, $\theta=\pi/2$, the stationary-limit surface satisfies
\begin{equation}
H_i(R_{\rm s})=0.
\label{eq:equatorial_static_limit_dimensionless}
\end{equation}
Therefore, the maximum separation between the event horizon and the stationary-limit surface occurs near the equatorial plane. The ergoregion is the domain
\begin{equation}
R_+ < R < R_{\rm s}(\theta),
\label{eq:ergoregion_domain_dimensionless}
\end{equation}
where the Killing vector $\partial_T$ becomes spacelike.

\begin{figure*}
\centering
\includegraphics[width=\textwidth]{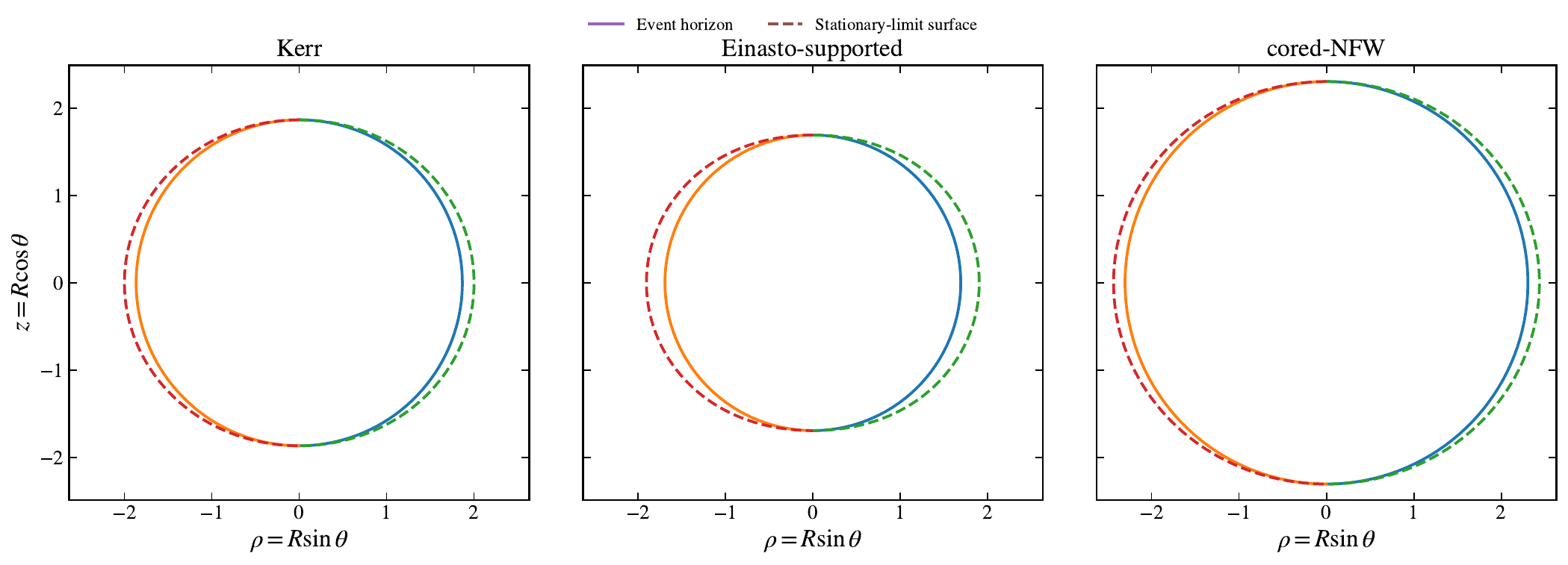}
\caption{
Meridional sections of the event horizon and stationary-limit surface for the Kerr reference spacetime and for representative rotating dark-matter-dressed geometries. The coordinates are $\rho=R\sin\theta$ and $z=R\cos\theta$. The region between the outer horizon and the stationary-limit surface defines the ergoregion. The parameter values are chosen within the black hole domain.
}
\label{fig:ergoregions}
\end{figure*}

A meridional representation of the horizon and stationary-limit surface is shown in Fig.~\ref{fig:ergoregions}. The coordinates used in the plot are $\rho=R\sin\theta$ and $z=R\cos\theta$. This figure illustrates how the dark matter dressing can slightly modify the near-horizon rotational structure before the ray  tracing analysis is performed. In the main imaging calculations, we restrict the parameter choices to the black hole region identified from the extremal condition above.

In practice, for each choice of dark matter profile and parameter set, we first solve Eq.~\eqref{eq:horizon_condition_dimensionless} to identify the existence and position of the outer horizon. We then solve Eq.~\eqref{eq:ergosurface_condition_3_dimensionless} for the stationary-limit surface. This preliminary classification is required before ray tracing, because the optical appearance can change qualitatively when the geometry approaches the extremal boundary or when the horizon disappears.

%%%%%%%%%%%%%%%%%%%%%%%%%%%%%%%%%%%%% Sect.3
\section{Null geodesics and observer-screen construction}
\label{sec:geodesics}

In this section, we describe the null-geodesic system used for the ray  tracing analysis. The rotating dark-matter-dressed geometries introduced in Section~\ref{sec:geometry} have a Kerr-like structure in which the profile dependence enters through the radial function $\Delta_i(R)$. For this particular ansatz, the Hamilton--Jacobi equation may admit
a Carter-type separation, similarly to the Kerr spacetime. However, we do not use this property in the numerical construction. Instead, we integrate the Hamiltonian equations directly, following the
standard phase-space formulation of null geodesic motion used in black hole ray tracing~\citep{Chandrasekhar1983,RauchBlandford1994,DexterAgol2009}. This choice keeps the ray  tracing pipeline independent of any
separability assumption and makes it applicable to more general rotating dark-matter-dressed geometries, for which the Carter-like
structure may be broken.

Throughout this section, the spacetime coordinates are denoted by
\begin{equation}
q^\mu=(T,R,\theta,\phi),
\label{eq:coordinate_vector_q}
\end{equation}
where $T=t/M$, $R=r/M$, and $\chi=a/M$ are the dimensionless quantities introduced in Section~\ref{sec:geometry}.

%%%%%%%%%%
\subsection{Hamiltonian formulation}
\label{subsec:hamiltonian}

Photon trajectories are described by the null condition
\begin{equation}
g_{\mu\nu}\dot{q}^{\mu}\dot{q}^{\nu}=0,
\label{eq:null_condition_velocity_dimensionless}
\end{equation}
where a dot denotes differentiation with respect to an affine parameter $\lambda$. Equivalently, one may introduce the Hamiltonian
\begin{equation}
\mathscr{H}
=
\frac{1}{2}g^{\mu\nu}p_{\mu}p_{\nu}
=
0,
\label{eq:null_hamiltonian_dimensionless}
\end{equation}
where $p_\mu$ is the photon momentum conjugate to $q^\mu$. The Hamiltonian equations are then
\begin{equation}
\frac{dq^\mu}{d\lambda}
=
\frac{\partial \mathscr{H}}{\partial p_\mu}
=
g^{\mu\nu}p_\nu,
\label{eq:hamilton_eq_q}
\end{equation}
and
\begin{equation}
\frac{dp_\mu}{d\lambda}
=
-\frac{\partial \mathscr{H}}{\partial q^\mu}
=
-\frac{1}{2}
\frac{\partial g^{\alpha\beta}}{\partial q^\mu}
p_\alpha p_\beta .
\label{eq:hamilton_eq_p_dimensionless}
\end{equation}
This formulation is convenient for numerical backward ray tracing because it only requires the inverse metric and its derivatives, and it does not require the existence of a separability structure beyond the conserved quantities associated with stationarity and axisymmetry \citep{Chandrasekhar1983,DexterAgol2009}.

The metric \eqref{eq:rotating_metric_dimensionless} is stationary and axisymmetric. Therefore, the photon energy and azimuthal angular momentum,
\begin{equation}
\mathcal{E}=-p_T,
\qquad
\ell=p_\phi,
\label{eq:energy_angular_momentum_dimensionless}
\end{equation}
are conserved along each ray. Hence, only $R$, $\theta$, $p_R$, and $p_\theta$ need to be evolved dynamically, while $p_T$ and $p_\phi$ are fixed by the initial conditions on the observer screen.

For the rotating metric \eqref{eq:rotating_metric_dimensionless}, the nonvanishing components of the inverse metric are
\begin{align}
g^{TT}
&=
-\frac{(R^2+\chi^2)^2-\chi^2\Delta_i(R)\sin^2\theta}
{\Sigma\,\Delta_i(R)},
\label{eq:ginv_TT_dimensionless}
\\
g^{T\phi}
&=
-\frac{\chi\,\mathcal{F}_i(R)}
{\Sigma\,\Delta_i(R)},
\label{eq:ginv_Tphi_dimensionless}
\\
g^{RR}
&=
\frac{\Delta_i(R)}{\Sigma},
\label{eq:ginv_RR_dimensionless}
\\
g^{\theta\theta}
&=
\frac{1}{\Sigma},
\label{eq:ginv_thetatheta_dimensionless}
\\
g^{\phi\phi}
&=
\frac{\Delta_i(R)-\chi^2\sin^2\theta}
{\Sigma\,\Delta_i(R)\sin^2\theta}.
\label{eq:ginv_phiphi_dimensionless}
\end{align}
Using these expressions, the null Hamiltonian becomes
\begin{align}
2\mathscr{H}
={}&
g^{TT}p_T^2
+2g^{T\phi}p_Tp_\phi
+g^{\phi\phi}p_\phi^2
+g^{RR}p_R^2
+g^{\theta\theta}p_\theta^2
=0 .
\label{eq:explicit_hamiltonian_dimensionless}
\end{align}
It is useful to note that the specific Kerr-like form adopted in
Eq.~\eqref{eq:rotating_metric_dimensionless} possesses a formal separability
structure. Indeed, using the Hamilton--Jacobi ansatz
\begin{equation}
S=-E T+L_z\phi+S_R(R)+S_\theta(\theta),
\end{equation}
and multiplying the null Hamilton--Jacobi equation by $\Sigma$, one obtains
\begin{multline}
\Delta_i(R)\left(\frac{dS_R}{dR}\right)^2
-\frac{\left[(R^2+\chi^2)E-\chi L_z\right]^2}{\Delta_i(R)}\\
+\left(\frac{dS_\theta}{d\theta}\right)^2
+\left(\frac{L_z}{\sin\theta}-\chi E\sin\theta\right)^2=0 .
\end{multline}
The first two terms depend only on $R$, whereas the last two terms
depend only on $\theta$. Therefore, for the present ansatz, a
Carter-like separation constant can be introduced. Nevertheless, in
the ray  tracing calculations below we retain the full Hamiltonian
system. This is numerically more direct and also avoids building the
method around a property that may not survive in more general
rotating dark-matter-dressed spacetimes.

The numerical integration is performed by evolving
\begin{align}
\dot{T}
&=
g^{TT}p_T+g^{T\phi}p_\phi,
\label{eq:Tdot_dimensionless}
\\
\dot{\phi}
&=
g^{T\phi}p_T+g^{\phi\phi}p_\phi,
\label{eq:phidot_dimensionless}
\\
\dot{R}
&=
\frac{\Delta_i(R)}{\Sigma}p_R,
\label{eq:Rdot_dimensionless}
\\
\dot{\theta}
&=
\frac{p_\theta}{\Sigma},
\label{eq:thetadot_dimensionless}
\end{align}
together with
\begin{equation}
\dot{p}_R
=
-\frac{1}{2}
\frac{\partial g^{\alpha\beta}}{\partial R}
p_\alpha p_\beta,
\label{eq:pRdot_dimensionless}
\end{equation}
and
\begin{equation}
\dot{p}_\theta
=
-\frac{1}{2}
\frac{\partial g^{\alpha\beta}}{\partial \theta}
p_\alpha p_\beta .
\label{eq:pthetadot_dimensionless}
\end{equation}
The null constraint~\eqref{eq:explicit_hamiltonian_dimensionless} is monitored along
the integration as a numerical consistency check. In the Kerr limit,
this formulation is equivalent to the usual null-geodesic system
written in terms of $E$, $L_z$, and the Carter constant
\citep{Bardeen1973,Chandrasekhar1983}. For the present
dark-matter-dressed ansatz, a Carter-like constant can also be
defined formally, as discussed above. However, the numerical
ray  tracing procedure is implemented through the full Hamiltonian
system, rather than through separated first-order equations. This
keeps the method robust and directly extensible to future rotating
dark-matter-dressed geometries where separability is not guaranteed.

%%%%%%%%%%
\subsection{Observer tetrad and image-plane coordinates}
\label{subsec:observer_screen}

The observer is placed at a large dimensionless radius $R_{\rm o}$ and inclination angle $\theta_{\rm o}$ with respect to the rotation axis. The observer screen is defined locally in the orthonormal frame of a stationary zero-angular-momentum observer (ZAMO) following the locally non-rotating frame construction commonly used in rotating black hole spacetimes \citep{BardeenPressTeukolsky1972,Chandrasekhar1983}. This choice provides a direct way of launching photons backward from the image plane toward the compact object.

For a stationary axisymmetric metric, the angular velocity of the local ZAMO is
\begin{equation}
\omega_{\rm o}
=
-\frac{g_{T\phi}}{g_{\phi\phi}}
\bigg|_{(R_{\rm o},\theta_{\rm o})}.
\label{eq:zamo_omega_dimensionless}
\end{equation}
The corresponding four-velocity is
\begin{equation}
u_{\rm o}^{\mu}
=
\mathcal{N}_{\rm o}^{-1}
\left(
1,0,0,\omega_{\rm o}
\right),
\label{eq:zamo_velocity_dimensionless}
\end{equation}
where the normalization factor is
\begin{equation}
\mathcal{N}_{\rm o}
=
\left[
-\left(
g_{TT}
+2\omega_{\rm o}g_{T\phi}
+\omega_{\rm o}^{2}g_{\phi\phi}
\right)
\right]^{1/2}_{(R_{\rm o},\theta_{\rm o})}.
\label{eq:zamo_normalization_dimensionless}
\end{equation}
A convenient orthonormal spatial triad is
\begin{align}
e_{(R)}^{\mu}
&=
\left(
0,\frac{1}{\sqrt{g_{RR}}},0,0
\right)_{(R_{\rm o},\theta_{\rm o})},
\label{eq:tetrad_eR_dimensionless}
\\
e_{(\theta)}^{\mu}
&=
\left(
0,0,\frac{1}{\sqrt{g_{\theta\theta}}},0
\right)_{(R_{\rm o},\theta_{\rm o})},
\label{eq:tetrad_etheta_dimensionless}
\\
e_{(\phi)}^{\mu}
&=
\left(
0,0,0,\frac{1}{\sqrt{g_{\phi\phi}}}
\right)_{(R_{\rm o},\theta_{\rm o})}.
\label{eq:tetrad_ephi_dimensionless}
\end{align}
Thus, the local tetrad is given by
\begin{equation}
\left\{
u_{\rm o}^{\mu},
e_{(R)}^{\mu},
e_{(\theta)}^{\mu},
e_{(\phi)}^{\mu}
\right\}.
\label{eq:observer_tetrad_set}
\end{equation}
By construction, this tetrad satisfies
$g_{\mu\nu}u_{\rm o}^{\mu}u_{\rm o}^{\nu}=-1$,
$g_{\mu\nu}e_{(a)}^{\mu}e_{(b)}^{\nu}=\delta_{ab}$, and
$g_{\mu\nu}u_{\rm o}^{\mu}e_{(a)}^{\nu}=0$. The choice
$\omega_{\rm o}=-g_{T\phi}/g_{\phi\phi}$ ensures that the azimuthal basis vector is orthogonal to the observer four-velocity.

We denote the Cartesian coordinates on the observer screen by $(X,Y)$, where $X$ measures the horizontal displacement and $Y$ the vertical displacement on the image plane. For a backward ray shot from the observer toward the black hole, the spatial direction in the local tetrad is chosen as
\begin{align}
n^{(R)}
&=
-\frac{R_{\rm o}}{\sqrt{R_{\rm o}^2+X^2+Y^2}},
\label{eq:nR_screen_dimensionless}
\\
n^{(\theta)}
&=
-\frac{Y}{\sqrt{R_{\rm o}^2+X^2+Y^2}},
\label{eq:ntheta_screen_dimensionless}
\\
n^{(\phi)}
&=
-\frac{X}{\sqrt{R_{\rm o}^2+X^2+Y^2}},
\label{eq:nphi_screen_dimensionless}
\end{align}
where $R_{\rm o}$ is a focal distance in the local orthonormal frame. The minus sign in $n^{(R)}$ ensures that the photon is initially directed inward. Hence, the unnormalised spatial direction associated with the screen point $(X,Y)$ is
$-R_{\rm o}e_{(R)}-Y e_{(\theta)}-X e_{(\phi)}$.
The minus sign in the $e_{(\phi)}$ component is chosen so that $X$ matches the usual horizontal celestial coordinate in the Kerr limit.

The initial photon four-momentum is then constructed as
\begin{equation}
p_{\rm o}^{\mu}
=
\mathcal{E}_{\rm loc}
\left[
u_{\rm o}^{\mu}
+n^{(R)}e_{(R)}^{\mu}
+n^{(\theta)}e_{(\theta)}^{\mu}
+n^{(\phi)}e_{(\phi)}^{\mu}
\right],
\label{eq:initial_momentum_tetrad_dimensionless}
\end{equation}
where $\mathcal{E}_{\rm loc}$ is the photon energy measured in the observer frame. Since null geodesics are invariant under an overall rescaling of the affine parameter, we set $\mathcal{E}_{\rm loc}=1$ without loss of generality. Since the spatial direction satisfies
$(n^{(R)})^2+(n^{(\theta)})^2+(n^{(\phi)})^2=1$,
the momentum in Eq.~\eqref{eq:initial_momentum_tetrad_dimensionless} is automatically null. 

Note that, although the ray is launched inward from the observer screen, the same null curve represents the path of a photon received by the observer after reversing the affine orientation. This standard backward ray tracing convention is used only to determine the geodesic path and its intersections with the emitting region.

The covariant components are obtained from
\begin{equation}
p_\mu=g_{\mu\nu}p^\nu
\label{eq:covariant_momentum_from_tetrad}
\end{equation}
at $(R_{\rm o},\theta_{\rm o})$, and the conserved quantities $p_T$ and $p_\phi$ are then fixed by the initial data.

In the limit of a distant observer in an asymptotically flat region, the screen coordinates $(X,Y)$ coincide with the usual celestial coordinates used to describe black hole shadows and ray-traced images of Kerr black holes \citep{Bardeen1973,CunninghamBardeen1972,Cunningham1975}. At finite but large $R_{\rm o}$, the tetrad construction avoids ambiguities associated with the coordinate basis and provides a well-defined local image plane for the numerical integration.

%%%%%%%%%%
\subsection{Capture, escape, and disk-intersection criteria}
\label{subsec:ray_classification}

Each ray is integrated backward from the observer screen until it satisfies one of the stopping criteria. A ray is classified as captured if it reaches the outer horizon,
\begin{equation}
R\leq R_{+}+\epsilon_{\rm h},
\label{eq:capture_condition_dimensionless}
\end{equation}
where $R_+$ is the outer event horizon obtained from Eq.~\eqref{eq:horizon_condition_dimensionless}, and $\epsilon_{\rm h}$ is a small numerical tolerance. A ray is classified as escaping if it reaches a large outer radius $R_{\rm max}$ with outward radial momentum after approaching the compact object. The boundary between captured and escaping rays defines the shadow contour on the observer screen.

For disk images, we also record intersections with the equatorial plane,
\begin{equation}
\theta=\frac{\pi}{2}.
\label{eq:equatorial_plane_dimensionless}
\end{equation}
In practice, an equatorial crossing is detected when the sign of $\theta-\pi/2$ changes between two consecutive integration steps. The crossing radius is then obtained by interpolation. If the crossing radius lies inside the emitting region,
\begin{equation}
R_{\rm in}\leq R_{\rm em}\leq R_{\rm out},
\label{eq:disk_region_dimensionless}
\end{equation}
the corresponding ray contributes to the observed disk image. The number of equatorial crossings provides a useful classification of the image order. The direct image corresponds to the first intersection with the disk, while higher-order crossings are associated with lensed images and photon-ring-related contributions, following the standard classification used in black hole lensing and photon ring studies \citep{CunninghamBardeen1972,GrallaHolzWald2019,Johnson2020}.

We denote by $N_{\rm cross}$ the number of equatorial intersections associated with a given ray. In the image-plane analysis, regions with different values of $N_{\rm cross}$ will be used to identify lensing bands and photon ring zones. This classification is especially useful for comparing the Kerr reference case with the rotating dark-matter-dressed geometries, because changes in the higher-order image structure can be more sensitive to the spacetime geometry than the primary shadow size alone. The construction described in this section defines the initial-value problem for each point of the observer screen. In the next section, this screen will be refined adaptively in order to resolve the shadow boundary, the lensing bands, and the photon ring region without requiring an excessively dense uniform grid. The same ray-classification criteria introduced above will then be used to separate the direct image from the higher-order contributions.

%%%%%%%%%%%%%%%%%%%%%%%%%%%%%%%%%%%%% Sect.4
\section{Adaptive ray tracing and transfer maps}
\label{sec:adaptive_ray_tracing}

In this section, we describe the numerical ray  tracing procedure used to connect the rotating geometries introduced in Section~\ref{sec:geometry} with the observer's image plane. The main purpose is to construct the shadow boundary, identify the lensing bands associated with successive equatorial crossings, and obtain the direct transfer map from the observer screen to the equatorial emission region. These ingredients will later be used to generate synthetic images of the accretion flow.

The ray tracing is performed backward from the observer screen. Each pixel on the screen, labelled by the dimensionless celestial coordinates $X$ and $Y$, determines an initial photon four-momentum through the tetrad construction described in Section~\ref{sec:geodesics}. The corresponding null geodesic is then evolved by integrating the Hamiltonian system. This backward ray  tracing strategy follows the standard logic of black hole imaging calculations, where null geodesics are traced from a distant observer toward the emitting region or the horizon \citep{CunninghamBardeen1973,Luminet1979}. In the Kerr spacetime, the shadow boundary and photon-ring-related structure can be studied analytically through the separability of the null geodesic equations \citep{Bardeen1973,Chandrasekhar1983,GrallaLupsasca2020a,GrallaLupsasca2020b}. In the rotating dark-matter-dressed metrics considered here, however, we do not assume such separability. The lensing information is therefore obtained numerically, by tracking the final fate of each ray and its intersections with the equatorial plane.

Our strategy is inspired by the lensing-band and adaptive ray  tracing viewpoint used in recent studies of black hole photon rings \citep{GrallaHolzWald2019,Johnson2020,CardenasAvendano2023}. In particular, instead of treating the whole image plane as a uniform grid with a single resolution requirement, we use the image-plane classification itself to identify the shadow boundary, the direct image region, and the higher-order lensed regions. This approach is especially useful for the present problem, because the dark-matter-dressed geometries must be handled numerically while still retaining a clear image-order interpretation.

%%%%%%%%%%
\subsection{Shadow-boundary tracing}
\label{subsec:shadow_boundary}

The first step is to determine the shadow boundary. Instead of constructing a very dense capture map on the whole image plane, we use a boundary-search method. For each polar direction on the observer screen,
\begin{equation}
X=b\cos\psi,
\qquad
Y=b\sin\psi,
\label{eq:screen_polar_boundary}
\end{equation}
we vary the impact parameter $b$ and locate the transition between captured and escaping rays. The shadow boundary is therefore obtained as
\begin{equation}
\mathcal{C}_{\rm sh}
=
\left\{
\left(X_{\rm sh}(\psi),Y_{\rm sh}(\psi)\right)
\right\},
\qquad
0\leq \psi <2\pi .
\label{eq:shadow_boundary_curve}
\end{equation}
This procedure is much more efficient than a uniform capture map, because only the transition curve is refined. It also provides a direct validation of the ray  tracing setup in the Kerr limit, where the critical curve is known analytically from the spherical photon orbits \citep{Bardeen1973,Chandrasekhar1983}.

\begin{figure}
    \centering
    \includegraphics[width=\columnwidth]{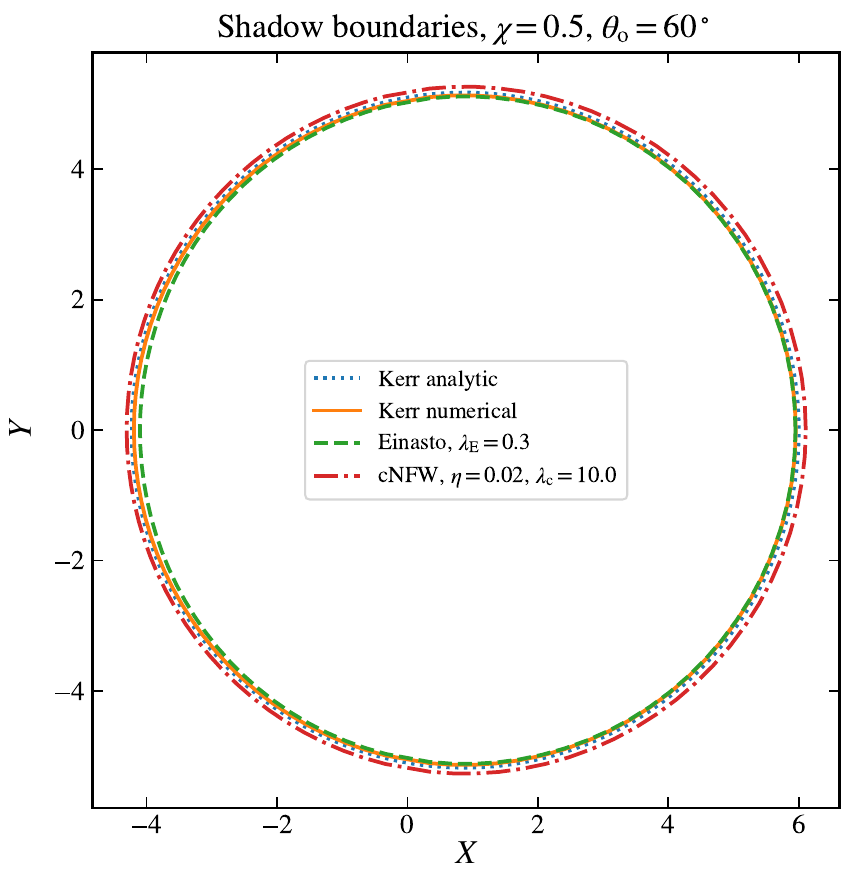}
    \caption{
    Shadow boundaries obtained by backward ray tracing for $\chi=0.5$ and $\theta_{\rm o}=60^\circ$. The dotted curve shows the analytical Kerr critical curve, while the solid Kerr curve is obtained numerically and provides a validation of the ray  tracing pipeline. The Einasto-supported and cored-NFW cases are shown for $\lambda_{\rm E}=0.3$, $\eta=0.02$, and $\lambda_{\rm c}=10.0$. 
    %The coordinates $X$ and $Y$ are dimensionless celestial coordinates on the observer screen.
    }
    \label{fig:shadow_boundaries}
\end{figure}

Figure~\ref{fig:shadow_boundaries} shows the resulting shadow boundaries for the Kerr reference spacetime, the rotating Einasto-supported black hole, and the rotating cored-NFW black hole. The dotted curve corresponds to the analytical Kerr critical curve, while the solid Kerr curve is obtained numerically with the same ray  tracing algorithm used for the dark-matter-dressed geometries. The agreement between these two curves validates the tetrad prescription, the screen convention, and the capture criterion. For the parameter values shown in the figure, the Einasto-supported model remains very close to the Kerr result, whereas the cored-NFW model gives a slightly larger apparent shadow. This behaviour indicates that, for the representative parameters adopted here, the modification induced by the cored-NFW profile is more visible in the primary shadow boundary than the Einasto correction.

%%%%%%%%%%
\subsection{Equatorial crossings and numerical lensing bands}
\label{subsec:lensing_bands}

To identify the image order, we count the number of intersections between each backward ray and the equatorial plane $\theta=\pi/2$. We denote this number by $N_{\rm cross}(X,Y)$. This construction is a numerical analogue of the image-order classification used in photon ring and lensing-ring studies, where higher-order images are associated with rays that undergo stronger deflection before reaching the observer \citep{GrallaHolzWald2019,Johnson2020,CardenasAvendano2023}.

The counting is performed before the ray either falls into the horizon or escapes back to the asymptotic region. In practice, the equatorial intersections are detected as events during the numerical integration of the geodesic equations. Near-tangent or spurious crossings are removed by imposing a minimum value of $|d\theta/d\lambda|$ at the crossing and by discarding repeated crossings that occur within a very small affine-parameter interval. This event-based procedure avoids interpreting numerical oscillations around the equatorial plane as physical image-order changes.

The numerical lensing bands are then defined by
\begin{equation}
\mathcal{B}_n
=
\left\{
(X,Y)\;|\; N_{\rm cross}(X,Y)\geq n+1
\right\},
\qquad n=0,1,2,\ldots .
\label{eq:lensing_bands_definition}
\end{equation}
With this convention, $\mathcal{B}_0$ corresponds to rays that intersect the equatorial plane at least once and therefore contribute to the direct image of an equatorial source. The band $\mathcal{B}_1$ contains rays that cross the equatorial plane at least twice and is associated with the first lensed image. Higher values of $N_{\rm cross}$ correspond to rays that spend more time in the strong-field region and approach the photon ring regime.

Equivalently, one may define the image layers
\begin{equation}
\mathcal{L}_n
=
\left\{
(X,Y)\;|\; N_{\rm cross}(X,Y)=n+1
\right\},
\label{eq:image_layers_definition}
\end{equation}
where $\mathcal{L}_0$ is the direct layer, $\mathcal{L}_1$ is the first lensed layer, and $\mathcal{L}_2$ is already associated with higher-order photon-ring-related contributions. This classification is purely numerical and therefore does not rely on the analytical separability of the Kerr geodesic equations.

\begin{figure*}
    \centering
    \includegraphics[width=\textwidth]{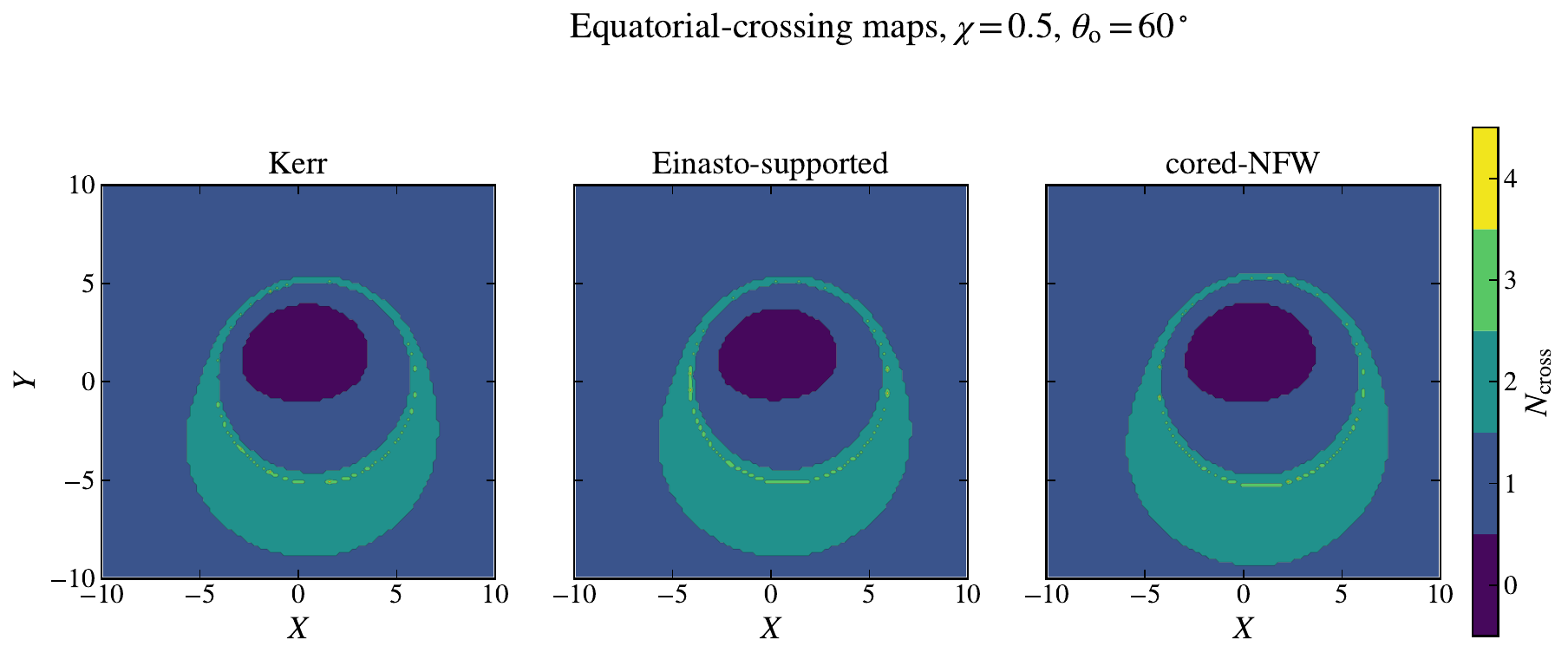}
    \caption{
    Equatorial-crossing maps $N_{\rm cross}(X,Y)$ for the Kerr, Einasto-supported, and cored-NFW geometries, with $\chi=0.5$ and $\theta_{\rm o}=60^\circ$. The colour indicates the number of trusted crossings of the equatorial plane before the ray is captured or escapes. The region with $N_{\rm cross}=0$ corresponds to rays that do not form a direct equatorial image, while $N_{\rm cross}=1$ and $N_{\rm cross}=2$ identify the direct and first lensed image domains, respectively.
    %The coordinates $X$ and $Y$ are dimensionless.
    }
    \label{fig:ncross_maps}
\end{figure*}

Figure~\ref{fig:ncross_maps} shows the equatorial-crossing maps for the three geometries. The central region with $N_{\rm cross}=0$ corresponds to rays that do not produce a direct equatorial image before capture. The surrounding region with $N_{\rm cross}=1$ is the direct-image domain, while the band with $N_{\rm cross}=2$ gives the first lensed contribution. The maps show that the general lensing-band structure is preserved in the dark-matter-dressed geometries, but the location and width of the bands are slightly modified. The cored-NFW case produces the most noticeable displacement of the lensing structure, consistently with the larger shadow boundary seen in Fig.~\ref{fig:shadow_boundaries}.

%%%%%%%%%%
\subsection{Direct transfer map}
\label{subsec:direct_transfer_map}

For each ray that intersects the equatorial plane, we record the crossing radius, azimuthal angle, and coordinate time. For the $n$th equatorial crossing, these quantities are denoted by
\begin{equation}
R_{\rm em}^{(n)}(X,Y),
\qquad
\phi_{\rm em}^{(n)}(X,Y),
\qquad
T_{\rm em}^{(n)}(X,Y).
\label{eq:transfer_functions}
\end{equation}
These functions define the numerical transfer map from the observer screen to the equatorial emission region. Similar transfer-function ideas have long been used in relativistic disk-imaging calculations, where the observed image is determined by the mapping between the local emitting matter and the observer screen, together with gravitational redshift, Doppler boosting, and lensing effects \citep{CunninghamBardeen1973,Luminet1979,GrallaHolzWald2019}. In the present work, the most important quantity at this stage is the direct transfer radius $R_{\rm em}^{(0)}(X,Y)$, which gives the radius of the first equatorial intersection.

The direct transfer map is useful for two reasons. First, it verifies that the direct image is being mapped smoothly from the observer screen to the source plane. Second, it provides the main geometrical input for the synthetic images constructed in the next section. For a prescribed equatorial emissivity profile $I_{\rm em}(R)$, the simplest direct-image intensity is controlled by $I_{\rm em}(R_{\rm em}^{(0)}(X,Y))$, before including additional redshift and optical-depth effects.

\begin{figure*}
    \centering    \includegraphics[width=\textwidth]{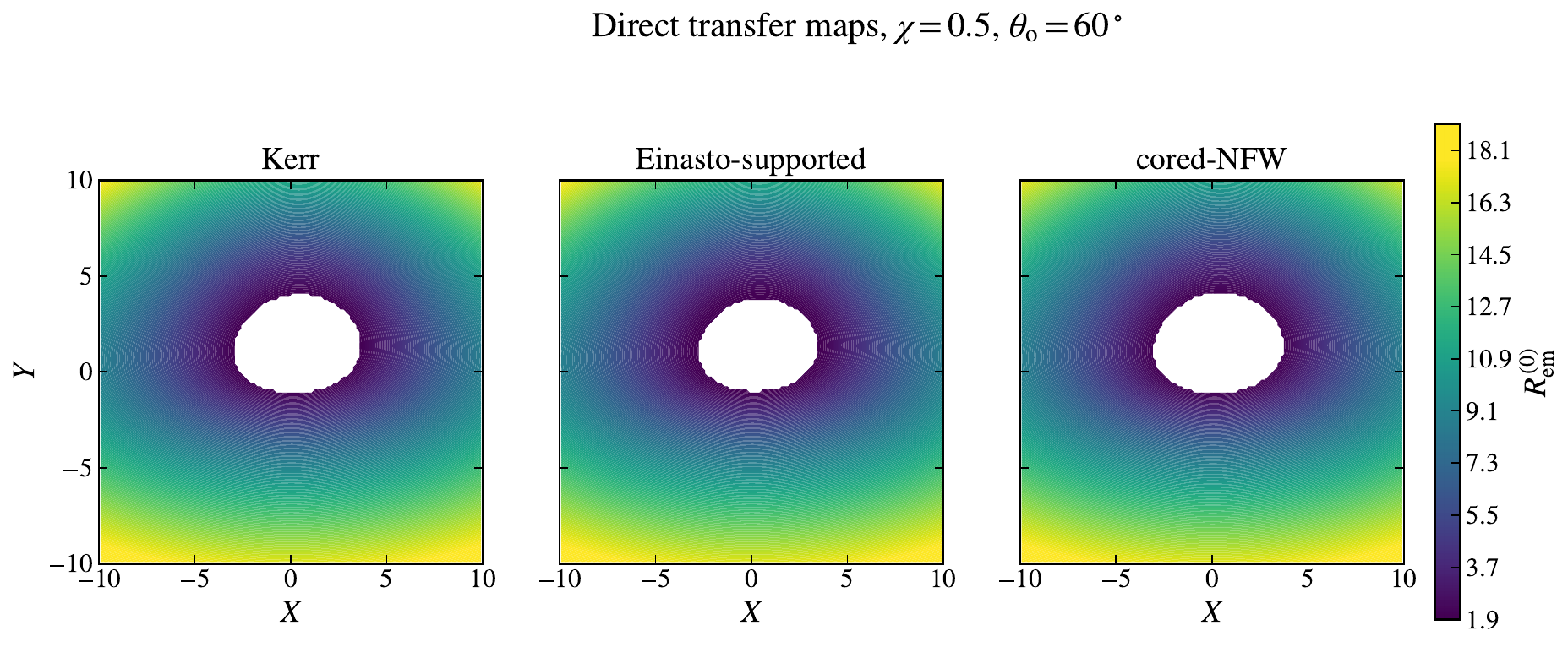}
    \caption{
    Direct transfer maps $R_{\rm em}^{(0)}(X,Y)$ for the Kerr, Einasto-supported, and cored-NFW geometries, with $\chi=0.5$ and $\theta_{\rm o}=60^\circ$. The colour represents the radius of the first trusted equatorial crossing of each backward ray. The white region corresponds to rays for which no direct equatorial crossing is recorded before capture or escape. These maps provide the geometrical input for the construction of direct synthetic images from an equatorial emissivity profile. 
    %The coordinates $X$ and $Y$ are dimensionless.
    }
    \label{fig:direct_transfer_maps}
\end{figure*}

Figure~\ref{fig:direct_transfer_maps} shows the direct transfer maps for the Kerr, Einasto-supported, and cored-NFW geometries. The white region corresponds to points on the observer screen for which no trusted first equatorial crossing is recorded before capture or escape. Outside this region, the colour indicates the radius of the first equatorial intersection. The maps are smooth and show that the direct image is reconstructed consistently in all three spacetimes. The small differences in the size and displacement of the central white region are consistent with the corresponding shadow-boundary shifts shown in Fig.~\ref{fig:shadow_boundaries}. These maps will be used in Section~\ref{sec:images} to construct the first synthetic accretion images.

%%%%%%%%%%
\subsection{Numerical strategy and role in the image construction}
\label{subsec:numerical_strategy_section4}

The results of this section provide the geometrical backbone of the imaging calculation. The shadow-boundary comparison validates the ray  tracing pipeline in the Kerr limit and shows the leading differences induced by the dark-matter-dressed geometries. The equatorial-crossing maps classify the image plane into direct and higher-order lensed regions. Finally, the direct transfer maps provide the source-plane radius associated with each pixel of the observer screen.

In the next section, we combine these transfer maps with semi-analytic emissivity prescriptions. This allows us to generate synthetic images without performing full GRMHD simulations. The aim is to isolate the effect of the background spacetime on the image morphology and photon-ring-related structure. More realistic time-dependent or GRMHD-inspired emission models can then be added as a further refinement once the geometrical ray  tracing framework has been validated.

%%%%%%%%%%%%%%%%%%%%%%%%%%%%%%%%%%%%% Sect.5
\section{Synthetic images from semi-analytic accretion models}
\label{sec:images}

The transfer maps constructed in Section~\ref{sec:adaptive_ray_tracing} provide the geometrical connection between the observer screen and the emitting region. In this section, we use these maps and the same backward ray  tracing framework to construct synthetic images of disk-like emission around the rotating geometries considered in this work. The aim is not to perform a full general-relativistic magnetohydrodynamical simulation, but rather to isolate the influence of the background spacetime on the image morphology through a controlled semi-analytic emission prescription.

This approach follows the standard logic of relativistic black hole imaging, where the observed image is obtained by combining null geodesic propagation with an emissivity model and the corresponding redshift factor \citep{CunninghamBardeen1973,Luminet1979,JaroszynskiKurpiewski1997,Dexter2016}. In the present case, we apply the same emission prescription to the Kerr, Einasto-supported, and cored-NFW geometries, so that differences in the resulting images can be attributed mainly to the change in the spacetime structure.

%%%%%%%%%%
\subsection{Semi-analytic disk emissivity}
\label{subsec:semi_analytic_emissivity}

We model the emitting matter as an optically thin, disk-like distribution concentrated around the equatorial plane. Similar phenomenological prescriptions are commonly used in ray-traced black hole imaging when the goal is to separate the geometrical lensing effect from the detailed plasma dynamics \citep{CunninghamBardeen1973,Luminet1979,Dexter2016}. The emissivity is taken to decrease radially and to be vertically suppressed away from the disk midplane. In terms of the cylindrical radius $\rho=R\sin\theta$ and vertical coordinate $z=R\cos\theta$, we use a phenomenological emissivity of the form
\begin{equation}
j(R,\theta)
=
j_0
\left(
\frac{\rho}{R_{\rm in}}
\right)^{-p}
\exp\left[
-\frac{\rho-R_{\rm in}}{R_{\rm s}}
\right]
\exp\left[
-\frac{z^2}{2H(\rho)^2}
\right]
\mathcal{T}_{\rm in}(\rho)
\mathcal{T}_{\rm out}(\rho),
\label{eq:emissivity_model}
\end{equation}
where $j_0$ is an arbitrary normalization, $p$ controls the radial fall-off, and $R_{\rm s}$ sets the radial emissivity scale. The function $H(\rho)$ specifies the effective vertical thickness of the emitting region. We use
\begin{equation}
H(\rho)=H_0+h\rho,
\label{eq:disk_height}
\end{equation}
where $H_0$ and $h$ are constant parameters. The functions $\mathcal{T}_{\rm in}$ and $\mathcal{T}_{\rm out}$, chosen as
\begin{align}
\mathcal{T}_{\rm in}(\rho)
&=
\frac{1}{2}
\left[
1+\tanh\left(\frac{\rho-R_{\rm in}}{\sigma_{\rm in}}\right)
\right],
\label{eq:taper_inner}
\\
\mathcal{T}_{\rm out}(\rho)
&=
\frac{1}{2}
\left[
1-\tanh\left(\frac{\rho-R_{\rm out}}{\sigma_{\rm out}}\right)
\right],
\label{eq:taper_outer}
\end{align}
are the smooth tapering functions that suppress the emission inside the inner disk edge and outside the outer disk radius. This avoids artificial discontinuities in the image. The smoothing scales $\sigma_{\rm in}$ and $\sigma_{\rm out}$ control
how rapidly the emission is switched on near the inner edge and
switched off near the outer edge. In the limit
$\sigma_{\rm in},\sigma_{\rm out}\rightarrow 0$, these functions approach
sharp step functions. In the present work they are kept finite only to
avoid artificial discontinuities in the image. 

The inner emitting radius is placed outside the event horizon according to
\begin{equation}
R_{\rm in}=R_+ + \Delta R_{\rm in},
\label{eq:rin_definition}
\end{equation}
where $R_+$ is the outer horizon radius of the corresponding rotating geometry. This choice ensures that the same prescription can be applied consistently to all three models, even when the horizon radius is shifted by the dark matter profile. This choice also allows the emitting region to
start close to the horizon; the taper only suppresses emission in the
unmodelled region below the prescribed inner edge, where the simple
circular-emitter prescription should not be over-interpreted.

We emphasize that Eq.~\eqref{eq:emissivity_model} is not intended to represent a unique physical plasma model. It is a controlled semi-analytic emissivity profile used to compare the optical appearance of the three backgrounds under the same emission assumptions.

%%%%%%%%%%
\subsection{Redshift-weighted observed intensity}
\label{subsec:redshift_intensity}

The observed intensity is computed by integrating the emissivity along each backward ray. In the optically thin approximation, the observed contribution from the emitting material is weighted by the redshift factor. Using the invariance of $I_\nu/\nu^3$ along null geodesics, which is the standard invariant form used in relativistic radiative transfer, the observed intensity is written schematically as \citep{CunninghamBardeen1973,JaroszynskiKurpiewski1997,Dexter2016}
\begin{equation}
I_{\rm obs}(X,Y)
\propto
\int_{\gamma(X,Y)}
g^3 j(R,\theta)\,d\lambda ,
\label{eq:observed_intensity_integral}
\end{equation}
where $\lambda$ is the affine parameter along the ray and $g$ is the redshift factor,
\begin{equation}
g
=
\frac{\nu_{\rm obs}}{\nu_{\rm em}}
=
\frac{-p_\mu u_{\rm obs}^{\mu}}
     {-p_\mu u_{\rm em}^{\mu}} .
\label{eq:redshift_factor}
\end{equation}
Here $p_\mu$ is the photon four-momentum, $u_{\rm obs}^{\mu}$ is the observer four-velocity, and $u_{\rm em}^{\mu}$ is the emitter four-velocity. For the observer tetrad used in this work, the photon energy measured by the distant observer is fixed by the initial normalization. The emitter is assumed to rotate approximately on circular equatorial orbits, with four-velocity
\begin{equation}
u_{\rm em}^{\mu}
=
u_{\rm em}^{T}
\left(
1,0,0,\Omega
\right),
\label{eq:emitter_four_velocity}
\end{equation}
where the angular velocity is obtained from the local metric functions as \citep{Bardeen1973,Chandrasekhar1983}
\begin{equation}
\Omega
=
\frac{
-\partial_R g_{T\phi}
+
\sqrt{
\left(\partial_R g_{T\phi}\right)^2
-
\left(\partial_R g_{TT}\right)
\left(\partial_R g_{\phi\phi}\right)
}
}
{\partial_R g_{\phi\phi}} .
\label{eq:keplerian_angular_velocity}
\end{equation}
The normalization $u_{\rm em}^{T}$ follows from $u_\mu u^\mu=-1$. This prescription captures the dominant gravitational-redshift and Doppler-boosting effects in a simple way, while avoiding the computational cost of a full plasma simulation.

Since the present images are intended as controlled geometrical comparisons, all models are rendered with the same emissivity parameters and the same display normalization. Therefore, the differences between the Kerr, Einasto-supported, and cored-NFW panels mainly reflect the different photon propagation and redshift structure of the corresponding spacetimes.

%%%%%%%%%%
\subsection{Synthetic images and image-order decomposition}
\label{subsec:synthetic_images_model_comparison}

The synthetic images are obtained by evaluating Eq.~\eqref{eq:observed_intensity_integral} for each pixel of the observer screen. The final intensity maps are displayed using an asinh stretch in order to reveal both the bright Doppler-enhanced region and the fainter lensed emission. Similar visualization choices are commonly used in black hole imaging because the intensity contrast between the direct and higher-order emission can be very large \citep{Luminet1979,Johnson2020,Dexter2016}. This stretch is used only for visualization and does not change the underlying ray-traced intensity data.

\begin{figure*}
    \centering
    \includegraphics[width=\textwidth]{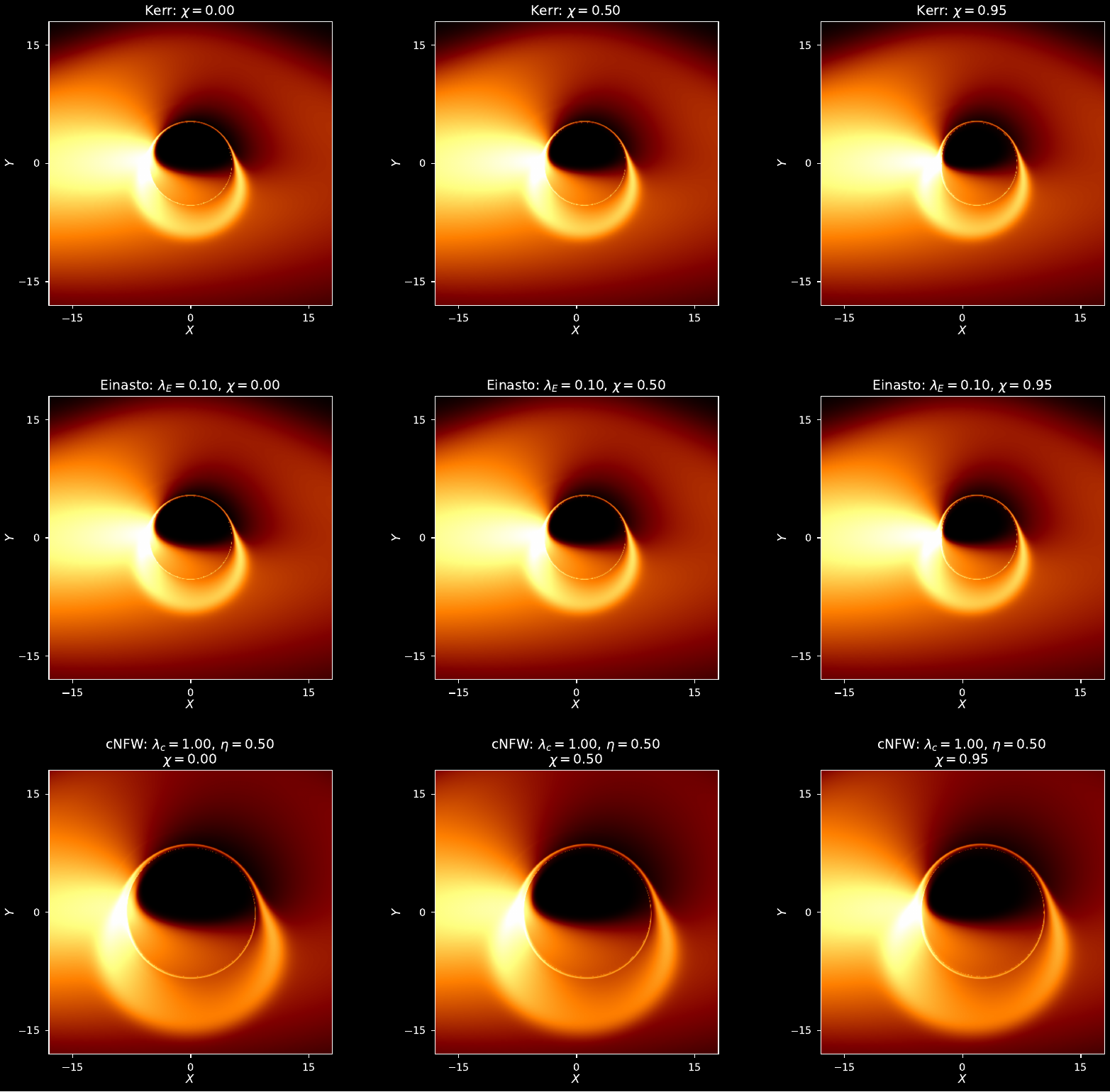}
    \caption{
    Synthetic images from the semi-analytic optically thin disk-like emission model at fixed inclination $\theta_{\rm o}=70^\circ$. The rows correspond to the Kerr, Einasto-supported, and cored-NFW geometries, while the columns show $\chi=0$, $\chi=0.5$, and $\chi=0.95$. The same emissivity prescription and display normalization are used in all panels. Increasing the spin enhances the image asymmetry and the Doppler-brightened side, while the cored-NFW model produces a larger apparent image scale for the representative parameters used here.
    }
    \label{fig:synthetic_spin_panel}
\end{figure*}

Figure~\ref{fig:synthetic_spin_panel} shows a spin sequence at fixed inclination $\theta_{\rm o}=70^\circ$. The rows correspond to the Kerr, Einasto-supported, and cored-NFW geometries, while the columns correspond to $\chi=0$, $\chi=0.5$, and $\chi=0.95$. For all three geometries, increasing the spin enhances the asymmetry of the image and shifts the bright Doppler-enhanced region. This behaviour is expected from frame dragging and from the stronger azimuthal velocity of the emitting material in the inner region. The Kerr and Einasto-supported images remain relatively close to each other for the adopted parameters, whereas the cored-NFW case produces a visibly larger apparent structure. This is consistent with the larger shadow boundary and modified transfer maps found in Section~\ref{sec:adaptive_ray_tracing}.

\begin{figure*}
    \centering
    \includegraphics[width=\textwidth]{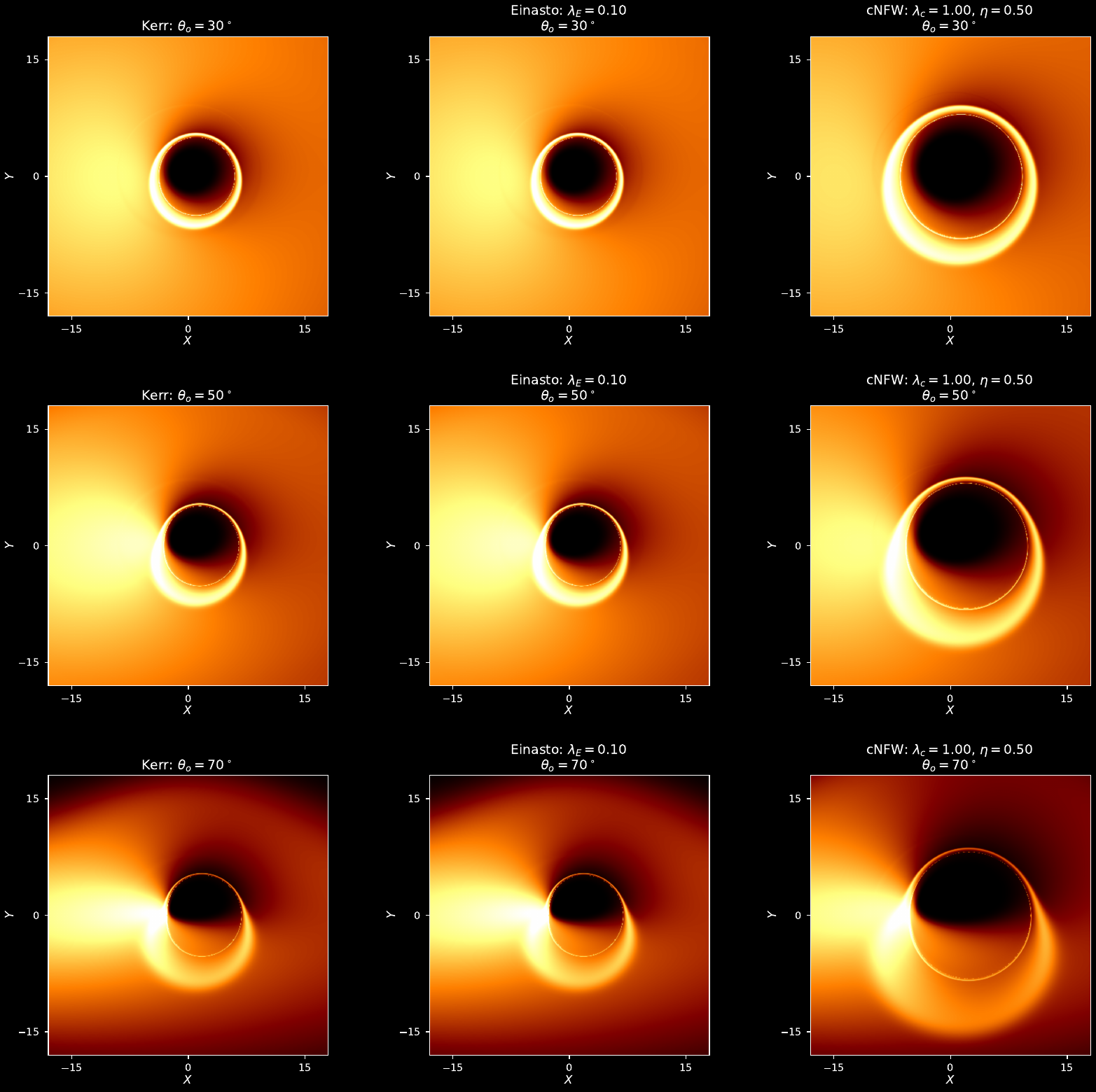}
    \caption{
    Synthetic images at fixed spin $\chi=0.95$ for different observer inclinations. The rows correspond to $\theta_{\rm o}=30^\circ$, $\theta_{\rm o}=50^\circ$, and $\theta_{\rm o}=70^\circ$, while the columns correspond to the Kerr, Einasto-supported, and cored-NFW geometries. Increasing the inclination makes the image more asymmetric and enhances the crescent-like morphology. The comparison illustrates how inclination and dark matter dressing can both affect the apparent size and brightness distribution of the image.
    }
    \label{fig:synthetic_inclination_panel}
\end{figure*}

Figure~\ref{fig:synthetic_inclination_panel} shows the complementary inclination sequence at fixed high spin $\chi=0.95$. The rows correspond to $\theta_{\rm o}=30^\circ$, $\theta_{\rm o}=50^\circ$, and $\theta_{\rm o}=70^\circ$, while the columns compare the Kerr, Einasto-supported, and cored-NFW geometries. As the inclination increases, the image becomes more asymmetric and the projected disk structure is increasingly distorted by relativistic beaming and gravitational lensing. The low-inclination images are comparatively more circular, while the high-inclination images display a stronger crescent-like morphology. Again, the cored-NFW case gives the largest apparent scale among the three models, while the Einasto-supported case remains closer to the Kerr reference for the parameter range considered here.

The two image sequences show that the dominant qualitative trends are controlled by spin and inclination, while the dark matter dressing introduces secondary but visible changes in the apparent image scale and morphology. In particular, for the selected parameters, the cored-NFW geometry produces the most pronounced deviation from the Kerr image, whereas the Einasto-supported case remains closer to Kerr. This result agrees with the behaviour observed in the shadow-boundary and transfer-map analyses. It also indicates that a possible degeneracy may arise between spin, inclination, and the dark matter parameters, since changes in the background geometry can partly mimic changes in the image size and asymmetry.

To separate the contribution of different lensed images, we also compute a representative image-order decomposition for the high-spin case $\chi=0.95$ and the inclination $\theta_{\rm o}=70^\circ$. In this diagnostic, the equatorial intersections of each backward ray are used to separate the direct image from the first and second lensed contributions. The direct image corresponds to the first disk intersection, while the first and second lensed images correspond to the next successive equatorial intersections. This construction follows the same image-order logic used in the lensing-band classification of Section~\ref{subsec:lensing_bands}, but now applied directly to the displayed intensity maps.

\begin{figure*}
    \centering
    \includegraphics[width=\textwidth]{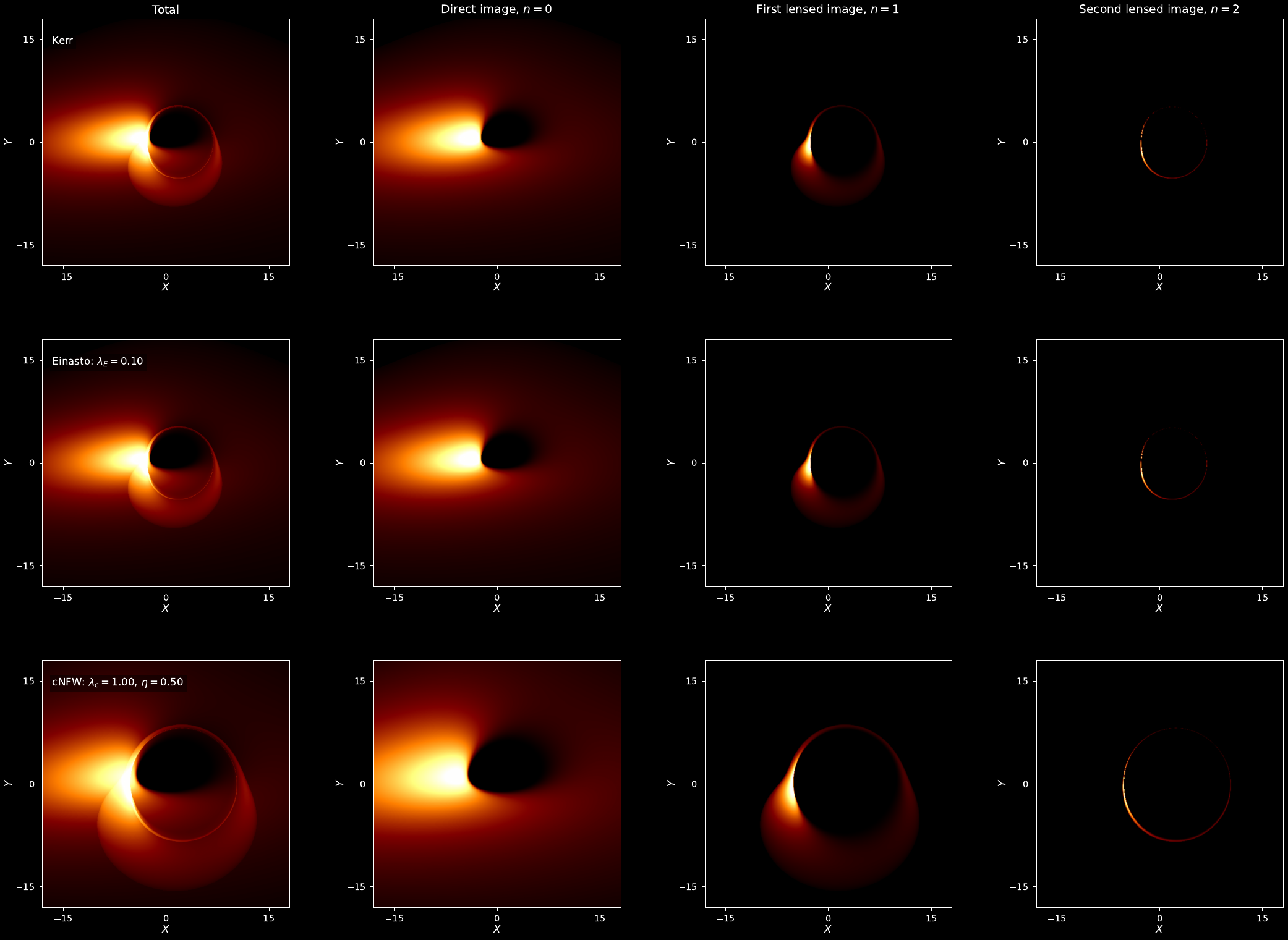}
    \caption{
    Representative image-order decomposition for $\chi=0.95$ and $\theta_{\rm o}=70^\circ$. The rows correspond to the Kerr, Einasto-supported, and cored-NFW geometries. The columns show the total image, the direct image $n=0$, the first lensed image $n=1$, and the second lensed image $n=2$. The direct image is associated with the first equatorial intersection of the backward ray, while the higher-order images correspond to subsequent equatorial intersections. Each panel is normalized independently in order to make the fainter higher-order contributions visible.
    }
    \label{fig:image_order_decomposition}
\end{figure*}

Figure~\ref{fig:image_order_decomposition} shows the resulting decomposition for the Kerr, Einasto-supported, and cored-NFW geometries. The first column gives the total image, while the remaining columns show the direct image, the first lensed image, and the second lensed image. The higher-order contributions are much fainter than the direct image in the underlying intensity data; therefore, the panels are normalized independently for visualization. This is only a display choice, made in order to show the morphology of the higher-order images. It should not be interpreted as implying that the second lensed image has a flux comparable to the direct image.

The decomposition confirms that the image-order structure remains qualitatively Kerr-like in all three cases, but its characteristic scale is modified by the dark matter dressing. The Einasto-supported geometry again stays close to the Kerr reference for the representative parameter value used here, while the cored-NFW case produces a larger and more displaced lensed structure. The second lensed image is narrow and ring-like, as expected for a higher-order contribution associated with rays that probe the strong-field region more closely. Therefore, the image-order decomposition supports the same conclusion obtained from the full synthetic images: the dark matter profile does not necessarily create a completely new image morphology, but it can change the apparent scale and the location of the lensed emission in a systematic way.

We emphasize that the present images should be interpreted as controlled ray  tracing experiments rather than predictions from a full accretion-flow simulation. The emissivity, disk thickness, and display stretch are fixed phenomenological choices. Nevertheless, applying the same prescription to all geometries allows us to isolate the role of the spacetime itself. A more realistic treatment would require polarized radiative transfer and time-dependent GRMHD emission models, which we leave for future work.

%%%%%%%%%%%%%%%%%%%%%%%%%%%%%%%%%%%%% Sect.6
\section{Phenomenological implications}
\label{sec:phenomenology}

In this section, we summarize the phenomenological meaning of the ray  tracing and synthetic-imaging results. The discussion is kept deliberately conservative. We do not fit EHT data and we do not claim that any of the geometries is favoured observationally. The aim is to identify which image-domain and Fourier-domain quantities are most sensitive to the dark matter dressing.

%%%%%%%%%%
\subsection{Image trends and spin--environment degeneracy}
\label{subsec:image_trends_degeneracy}

The results of Sections~\ref{sec:adaptive_ray_tracing} and \ref{sec:images} show a common Kerr-like structure in all three geometries. Increasing the observer inclination enhances the crescent-like asymmetry, while increasing the spin modifies the displacement and brightness contrast of the image. Against this common behaviour, the dark matter dressing introduces profile-dependent shifts in the apparent scale and in the location of the lensed emission.

For the representative parameters used here, the Einasto-supported geometry remains close to the Kerr reference. The cored-NFW geometry gives the larger departure, already visible in the shadow boundary of Fig.~\ref{fig:shadow_boundaries}, in the lensing-band and transfer-map structures of Figs.~\ref{fig:ncross_maps} and \ref{fig:direct_transfer_maps}, and in the synthetic images of Figs.~\ref{fig:synthetic_spin_panel} and \ref{fig:synthetic_inclination_panel}. This indicates a partial spin--environment degeneracy: changes in the dark matter profile can mimic part of the effect usually attributed to spin, inclination, or emission structure. In the present work this degeneracy is qualitative, not a statistical constraint, because the emission model is fixed and no parameter estimation is performed.

%%%%%%%%%%
\subsection{Image-level morphology diagnostics}
\label{subsec:morphology_diagnostics}

The image sequences discussed above show that the dark matter dressing can modify the apparent scale, displacement, and brightness structure of the ray-traced images. To quantify this behaviour, we use simple diagnostics computed from the raw intensity map $I(X,Y)$. Similar image-domain quantities, such as characteristic diameter, ring width, centroid displacement, and brightness asymmetry, are commonly used in the interpretation of horizon-scale black hole images and interferometric reconstructions \citep{Chael2018,EHT2019IV,EHT2019VI,EHT2022IV}. Here they are not meant to replace a visibility-domain analysis, but to provide a compact comparison among the Kerr, Einasto-supported, and cored-NFW images under the same emission prescription.

For a given image, we define the total flux
\begin{equation}
F=\int I(X,Y)\,dX\,dY .
\end{equation}
The intensity-weighted centroid is
\begin{equation}
X_c=\frac{1}{F}\int X I(X,Y)\,dX\,dY,\qquad
Y_c=\frac{1}{F}\int Y I(X,Y)\,dX\,dY,
\end{equation}
and the centroid displacement is
\begin{equation}
C=\sqrt{X_c^2+Y_c^2}.
\end{equation}
We then define the centroid-centred radius
\begin{equation}
\varrho(X,Y)=\sqrt{(X-X_c)^2+(Y-Y_c)^2},
\end{equation}
the intensity-weighted mean radius
\begin{equation}
\bar{\varrho}=\frac{1}{F}\int \varrho(X,Y) I(X,Y)\,dX\,dY,
\end{equation}
and the effective image diameter
\begin{equation}
d_{\rm img}=2\bar{\varrho}.
\end{equation}
The radial width is measured by
\begin{equation}
w_{\rm img}=\left[\frac{1}{F}\int
\left(\varrho-\bar{\varrho}\right)^2 I(X,Y)\,dX\,dY
\right]^{1/2},
\end{equation}
so that $w_{\rm img}/d_{\rm img}$ gives a dimensionless measure of the relative image thickness.

The left--right brightness asymmetry is defined with respect to the centroid as
\begin{equation}
A_{\rm LR}=\frac{F_R-F_L}{F_R+F_L},
\end{equation}
where
\begin{equation}
F_R=\int_{X>X_c} I(X,Y)\,dX\,dY,\qquad
F_L=\int_{X<X_c} I(X,Y)\,dX\,dY .
\end{equation}
Its sign depends on the orientation convention, while $|A_{\rm LR}|$ measures the strength of the crescent asymmetry. We also use the peak-to-mean contrast
\begin{equation}
\mathcal{P}=\frac{I_{\rm max}}{\langle I\rangle_{\rm img}},
\end{equation}
where $\langle I\rangle_{\rm img}$ is the mean intensity over the emitting image region. Finally, for comparison with Kerr we define
\begin{equation}
\delta Q_i=\frac{Q_i-Q_{\rm Kerr}}{Q_{\rm Kerr}},
\end{equation}
where $Q_i$ denotes any of the above diagnostics for the dark-matter-dressed model. All values below are computed from the raw ray-traced intensity maps, not from the asinh-stretched display images.

\begin{table*}
\centering
\caption{Image-level morphology diagnostics for the Kerr,
Einasto-supported, and cored-NFW geometries. The quantities are
computed from the raw ray-traced intensity maps, not from the
asinh-stretched display images. Here $d_{\rm img}$ is the effective
image diameter, $C$ is the centroid displacement, $w_{\rm img}/d_{\rm img}$
is the relative image width, $|A_{\rm LR}|$ is the left--right brightness
asymmetry, $\mathcal{P}=I_{\rm max}/\langle I\rangle_{\rm img}$ is the
peak-to-mean contrast, and $\delta d_{\rm img}$ gives the relative
change of the effective image diameter with respect to Kerr.}
\label{tab:morphology_diagnostics}
\begin{tabular}{lcccccccc}
\hline
Model & $\chi$ & $\theta_o$ & $d_{\rm img}$ & $C$ &
$w_{\rm img}/d_{\rm img}$ & $|A_{\rm LR}|$ & $\mathcal{P}$ &
$\delta d_{\rm img}$ \\
\hline
Kerr      & 0.95 & $70^\circ$ & 16.907 & 3.652 & 0.270 & 0.192 & 9.666  & 0.000 \\
Einasto   & 0.95 & $70^\circ$ & 16.906 & 3.653 & 0.270 & 0.195 & 9.671  & 0.000 \\
cored-NFW & 0.95 & $70^\circ$ & 16.860 & 5.126 & 0.281 & 0.316 & 18.261 & -0.003 \\
\hline
\end{tabular}
\end{table*}

Table~\ref{tab:morphology_diagnostics} makes the image-level trend more explicit. The Einasto-supported image is almost indistinguishable from Kerr for the adopted value of $\lambda_{\rm E}$. The cored-NFW image has nearly the same intensity-weighted diameter, but a larger centroid displacement, stronger left--right asymmetry, and larger peak-to-mean contrast. Thus, in this representative case, the main cored-NFW effect is not simply a global rescaling of the image. It is a redistribution of the bright emission on the observer screen.

%%%%%%%%%%
\subsection{Angular-scale confrontation with M87* and Sgr A*}
\label{subsec:angular_size_confrontation}

The diagnostics above are dimensionless. To connect them with horizon-scale observations, a dimensionless image scale $d$ can be converted with the gravitational angular radius
\begin{equation}
\theta_g=\frac{G M}{c^2D},
\end{equation}
where $M$ and $D$ are the source mass and distance. The corresponding angular scale is
\begin{equation}
\theta=d\,\theta_g .
\end{equation}
For reference, the EHT measured a characteristic crescent diameter $\theta_d=42\pm3\,\mu{\rm as}$ for M87* and inferred $\theta_g=3.8\pm0.4\,\mu{\rm as}$ \citep{EHT2019VI}. For Sgr A*, the image-domain analysis found a characteristic ring diameter $\theta_d=51.8\pm2.3\,\mu{\rm as}$, with the gravitational angular scale fixed by the Galactic centre mass--distance calibration \citep{EHT2022IV,EHT2022VI}. These EHT values are used below only as reference angular scales.

Since $d_{\rm img}$ is an intensity-weighted measure of the full emitting image, it should not be identified directly with the EHT crescent diameter. We therefore define a second scale from the origin-centred radial intensity profile. With
\begin{equation}
\varrho_0=\sqrt{X^2+Y^2},
\end{equation}
we bin the raw intensity map as a function of $\varrho_0$ and locate the peak radius $\varrho_{\rm peak}$. The corresponding characteristic bright-peak scale is
\begin{equation}
d_{\rm peak}=2\varrho_{\rm peak}.
\end{equation}
This is not the mathematical shadow diameter, and it is not the EHT ring diameter obtained from an observational fit. It is a model-dependent bright-peak scale extracted from our synthetic image. Its angular version is
\begin{equation}
\theta_{\rm peak}=d_{\rm peak}\theta_g .
\end{equation}
This quantity gives a reproducible way of comparing the three synthetic images at the level of characteristic emission scale, while keeping the interpretation separate from a true EHT model fit.

\begin{figure*}
\centering
\begin{minipage}[t]{0.49\textwidth}
\centering
\vspace{0pt}
\includegraphics[width=\linewidth]{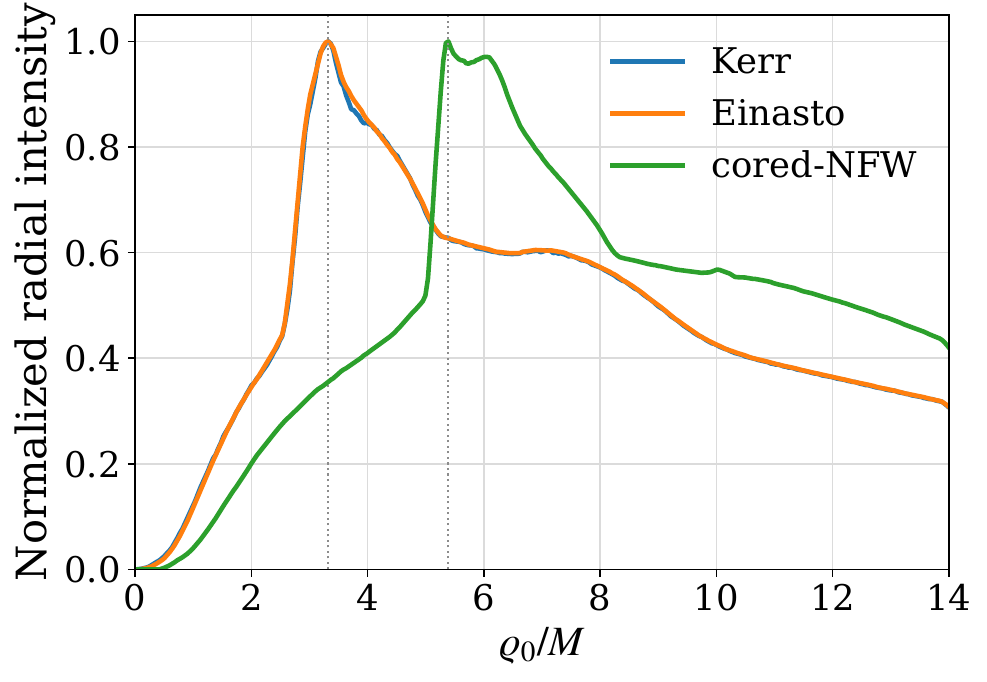}
\end{minipage}
\hfill
\begin{minipage}[t]{0.5\textwidth}
\centering
\vspace{0pt}
\includegraphics[width=1.02\linewidth]{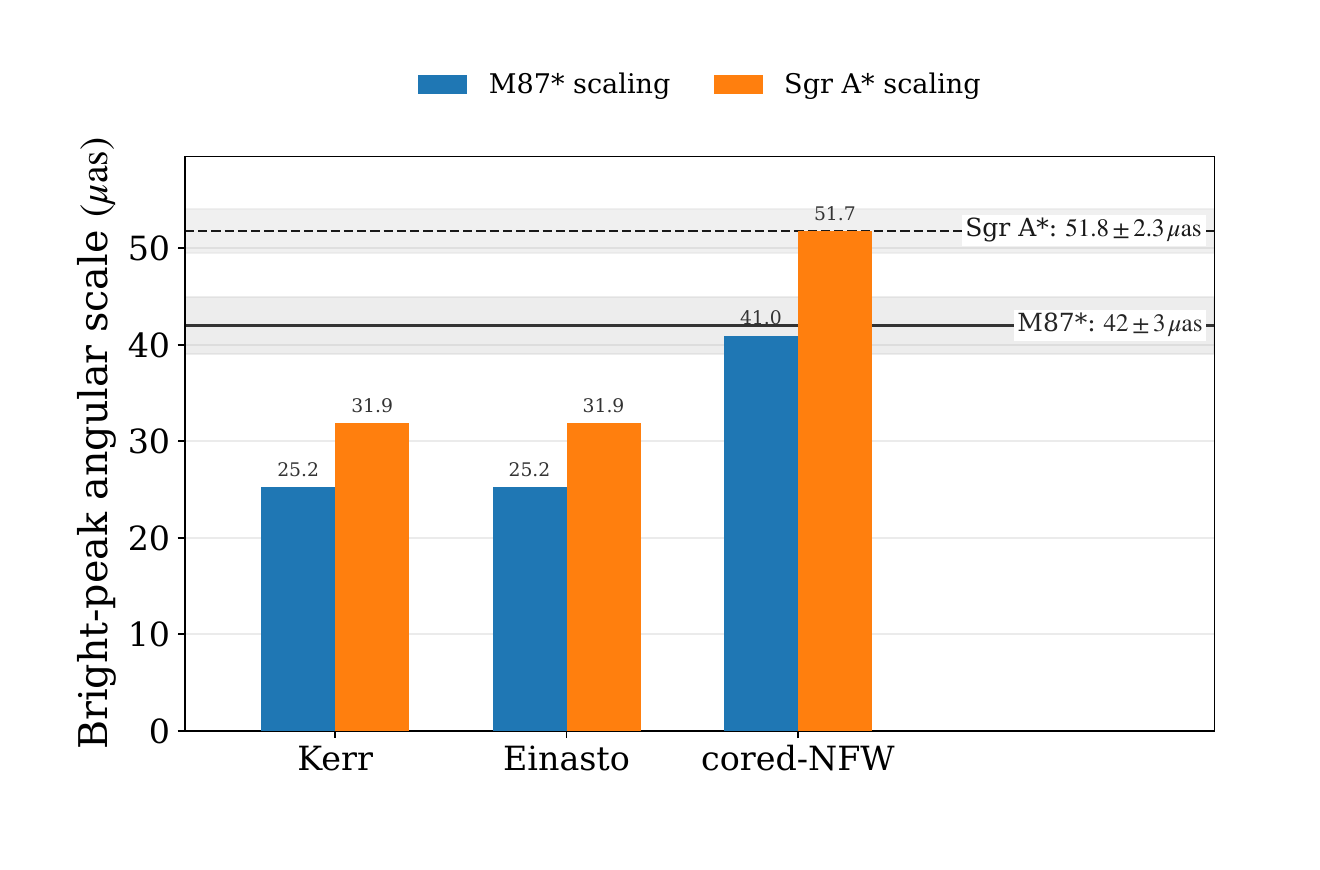}
\end{minipage}
\caption{Angular-scale confrontation for the representative high-spin and high-inclination images with $\chi=0.95$ and $\theta_o=70^\circ$. Left: origin-centered radial intensity profiles computed from the raw ray-traced images. The dotted vertical lines mark the peak radii used to define $d_{\rm peak}=2\varrho_{\rm peak}$. Right: bright-peak angular scales obtained from $\theta_{\rm peak}=d_{\rm peak}\theta_g$ using the gravitational angular scales of M87* and Sgr A*. The horizontal bands indicate the characteristic EHT ring-diameter reference ranges. These bands are included only as observational reference scales. The quantity $\theta_{\rm peak}$ is not the shadow diameter and is not obtained from an EHT image fit.}
\label{fig:angular_size_confrontation}
\end{figure*}

\begin{table*}
\centering
\caption{Angular-scale confrontation for the representative high-spin and high-inclination images. The dimensionless quantity $d_{\rm peak}$ is obtained from the peak of the origin-centred radial intensity profile of the raw ray-traced image. The corresponding bright-peak angular scales are computed using $\theta_{\rm peak}=d_{\rm peak}\theta_g$, with $\theta_g=3.8\,\mu{\rm as}$ for M87* and $\theta_g=4.8\,\mu{\rm as}$ for Sgr A*. The EHT reference ring diameters, $42\pm3\,\mu{\rm as}$ for M87* and $51.8\pm2.3\,\mu{\rm as}$ for Sgr A*, are shown only as reference angular scales, not as fitted quantities.}
\label{tab:angular_size_confrontation}
\begin{tabular}{lccccc}
\hline
Model & $\chi$ & $\theta_o$ & $d_{\rm peak}$ &
$\theta_{\rm peak}^{\rm M87*}\,(\mu{\rm as})$ &
$\theta_{\rm peak}^{\rm Sgr A*}\,(\mu{\rm as})$ \\
\hline
Kerr      & 0.95 & $70^\circ$ & 6.644  & 25.246 & 31.890 \\
Einasto   & 0.95 & $70^\circ$ & 6.644  & 25.246 & 31.890 \\
cored-NFW & 0.95 & $70^\circ$ & 10.779 & 40.962 & 51.741 \\
\hline
\end{tabular}
\end{table*}

Figure~\ref{fig:angular_size_confrontation} and Table~\ref{tab:angular_size_confrontation} show that Kerr and the Einasto-supported model have the same bright-peak scale for the representative parameters used here. In this particular semi-analytic emission model, their origin-centred radial intensity peaks lie at a smaller angular scale than the EHT reference ring diameters. This should not be interpreted as an exclusion of Kerr or of the Einasto-supported model, because $\theta_{\rm peak}$ is not a fitted EHT ring diameter and the comparison does not include a GRRT or GRMHD emission model, scattering, beam convolution, visibility-domain sampling, or marginalization over source and nuisance parameters. The cored-NFW profile shifts the radial peak outward, giving $d_{\rm peak}=10.779$. With the M87* angular scale this corresponds to $40.962\,\mu{\rm as}$, close to the reference diameter $42\pm3\,\mu{\rm as}$. With the Sgr A* angular scale it gives $51.741\,\mu{\rm as}$, close to $51.8\pm2.3\,\mu{\rm as}$. This is not evidence that the cored-NFW geometry is preferred by EHT data. It only shows that, under this simple angular conversion and this chosen emissivity prescription, the cored-NFW dressing can move the model-dependent bright-peak scale into an observationally relevant range.

%%%%%%%%%%
\subsection{Simplified visibility-amplitude diagnostics}
\label{subsec:visibility_diagnostics}

Very-long-baseline interferometers such as the EHT do not measure images directly. They sample complex visibilities, which are Fourier components of the sky brightness distribution \citep{Thompson2017,EHT2019IV,Chael2018}. To make a first connection with this language, we compute a simplified visibility-amplitude diagnostic from the synthetic images. This is not an EHT visibility-domain fit: no baseline coverage, noise, calibration, closure quantities, or scattering model is included.

For an image $I(X,Y)$ we define
\begin{equation}
V(u,v)=\int I(X,Y)\exp\left[-2\pi i(uX+vY)\right]dX\,dY,
\end{equation}
and normalize the amplitude by the zero-baseline flux,
\begin{equation}
\mathcal{V}(u,v)=\frac{|V(u,v)|}{|V(0,0)|}.
\end{equation}
We use the azimuthally averaged amplitude $\mathcal{V}(q)$, with $q=\sqrt{u^2+v^2}$, and the horizontal cut $\mathcal{V}(|u|,0)$ as compact Fourier-domain summaries.
\begin{figure*}
\centering
\includegraphics[width=\textwidth]{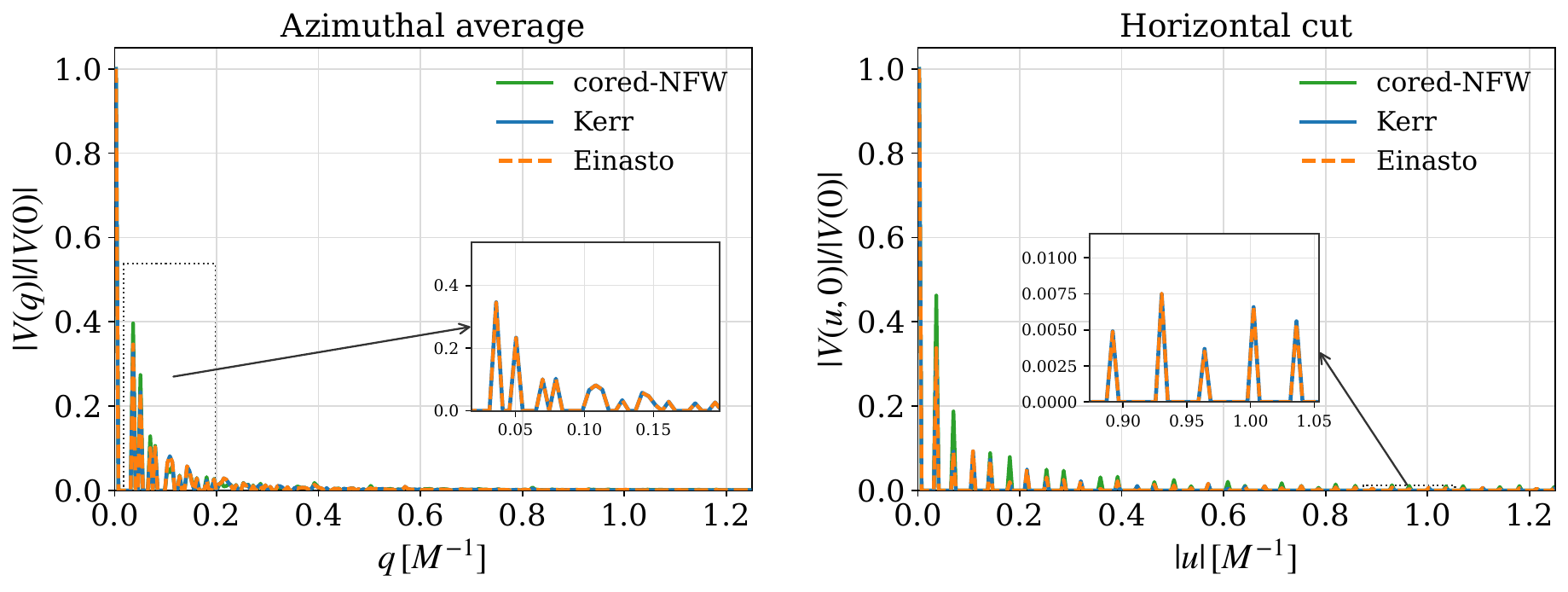}
\caption{Simplified visibility-amplitude diagnostics for the representative high-spin and high-inclination images with $\chi=0.95$ and $\theta_o=70^\circ$. The left panel shows the azimuthally averaged normalized visibility amplitude $|V(q)|/|V(0)|$, while the right panel shows the horizontal cut $|V(u,0)|/|V(0)|$. The insets show only the Kerr and Einasto-supported curves, in order to magnify their small separation; the cored-NFW curve is omitted from the insets because its departure is already visible in the main panels. These curves are computed from the raw synthetic images and are intended only as Fourier-domain diagnostics, not as an EHT visibility-domain fit.}
\label{fig:visibility_diagnostics}
\end{figure*}
Figure~\ref{fig:visibility_diagnostics} is consistent with the image-domain results. Kerr and the Einasto-supported model remain almost indistinguishable over the displayed Fourier scales, while the cored-NFW profile gives a visibly different normalized amplitude. This is expected because the cored-NFW image has a different characteristic bright-ring scale and a stronger redistribution of intensity. A real comparison with EHT data would require the source angular scaling, scattering treatment for Sgr A*, baseline sampling, instrumental response, calibration uncertainties, and nuisance-parameter marginalization \citep{EHT2019IV,EHT2019V,Johnson2020}.

%%%%%%%%%%
\subsection{Scope, limitations, and future directions}
\label{subsec:limitations_future}

The results should be read within the scope of a controlled phenomenological ray  tracing study. The rotating metrics are effective Kerr-like extensions of static dark-matter-dressed seed geometries. They are useful for isolating profile-dependent optical effects, but they are not presented as unique dynamical rotating solutions of the full matter sector. Likewise, the effective stress-energy tensor of the rotating spacetime is not used here to build a self-consistent accretion model.

The emission model is also deliberately simple. We use a stationary, optically thin, finite-thickness disk-like emissivity and apply it in the same way to Kerr, Einasto-supported, and cored-NFW geometries. This makes the comparison clean, but it does not replace GRRT calculations based on GRMHD accretion flows \citep{Dexter2016,EHT2019V,EHT2022V}. In particular, the present images do not include self-consistent plasma dynamics, magnetic-field structure, electron thermodynamics, polarization, absorption, Faraday effects, or turbulent variability.

For the same reason, the angular-scale and visibility comparisons with M87* and Sgr A* are confrontations, not constraints. A direct observational test would require mass--distance calibration, flux normalization, scattering treatment for Sgr A*, EHT-like beam convolution, realistic baseline coverage, visibility-domain or closure-quantity modelling, and marginalization over spin, inclination, emission parameters, and the dark matter profile parameters. Without such a pipeline, one cannot claim that a specific dark-matter-dressed geometry is favoured or excluded by present EHT observations.

The value of the present framework is that it identifies which quantities are worth following in a more expensive analysis. For the representative parameters considered here, the Einasto-supported model remains nearly degenerate with Kerr, whereas the cored-NFW model modifies the morphology diagnostics, the characteristic bright-ring scale, and the normalized visibility amplitude. This suggests that future work should explore the dark matter parameters jointly with spin, inclination, and emission prescriptions, and then replace the semi-analytic emissivity by GRRT or GRMHD-informed emission models. For Sgr A*, the scattering kernel should also be included before any direct comparison with reconstructed image scales.

%%%%%%%%%%%%%%%%%%%%%%%%%%%%%%%%%%%%% Sect.7
\section{Conclusions}
\label{sec:conclusions}

We have studied the optical appearance of rotating black holes dressed by dark matter profiles. Starting from Einasto-supported and cored-NFW static seed geometries, we constructed effective rotating extensions and compared them with Kerr using the same ray  tracing and emission prescription. The purpose was not to fit EHT data or to claim a dark matter detection, but to see how the dark matter dressing changes the shadow, lensing structure, synthetic image morphology, and simple Fourier-domain diagnostics.

The spacetime analysis showed that the dark matter parameters modify the horizon function, the extremal spin curve, and the ergoregion. These changes remain moderate for the representative cases considered here, but they provide the geometrical origin of the differences found later in the ray  tracing results. The Hamiltonian formulation allowed us to integrate the null geodesics directly, without relying on separated first-order equations. This keeps the numerical pipeline useful also for more general rotating backgrounds where separability may not survive.

The ray  tracing results were first validated in the Kerr limit, where the numerical shadow agrees with the analytical critical curve. We then applied the same method to the Einasto-supported and cored-NFW geometries. The Einasto-supported case stays close to Kerr for the adopted parameters, while the cored-NFW case produces the larger shift in the shadow boundary, lensing bands, and transfer maps. The synthetic images show the expected dependence on spin and inclination, but also reveal profile-dependent changes in the scale and brightness distribution of the lensed emission.

The morphology diagnostics make this point more quantitative. In the representative high-spin and high-inclination case, the Einasto-supported image is almost identical to Kerr. The cored-NFW image has a similar intensity-weighted diameter, but a larger centroid displacement, stronger brightness asymmetry, and higher peak-to-mean contrast. The origin-centred radial profile also shows an outward shift of the model-dependent bright-peak scale. When converted with the gravitational angular scales of M87* and Sgr A*, the cored-NFW bright-peak scale lies close to the corresponding EHT reference ring-diameter ranges. This agreement should not be interpreted as a fit or as a preference for cored-NFW, and the smaller Kerr and Einasto bright-peak scales should not be read as an exclusion of those models. It only shows that the cored-NFW effect is large enough to be observationally relevant at the level of this particular angular-scale diagnostic.

The simplified visibility-amplitude diagnostic gives the same qualitative picture. Kerr and the Einasto-supported model remain nearly degenerate, while the cored-NFW model produces a visible change in the normalized Fourier-domain amplitude. Thus, the dark matter dressing does not need to create a completely new image morphology in order to affect quantities that are relevant for horizon-scale imaging. It may instead enter through centroid shifts, brightness redistribution, bright-ring scale, and visibility-amplitude structure, all of which are also degenerate with spin, inclination, and emission physics.

A full confrontation with M87* or Sgr A* is left for future work. Such a comparison will require GRRT or GRMHD-informed emission models, scattering treatment for Sgr A*, EHT-like beam and baseline modelling, and marginalization over mass--distance calibration, flux normalization, spin, inclination, emission parameters, and dark matter profile parameters. The present framework provides the geometrical and diagnostic basis for that next step.

%%%%%%%%%%%%%%%%%%%%%%%%%%%%%%%%%%%%%%%%%%%%%%%%%%
\section*{Acknowledgments}

The author acknowledges financial support from Agencia Nacional de Investigaci\'on y Desarrollo (ANID) through the FONDECYT Postdoctoral Grant No.~3260029. The author also acknowledges that ChatGPT (OpenAI) was used only as an auxiliary tool for language polishing and improving the clarity and readability of the manuscript. All scientific developments, calculations, interpretations, figures, and final revisions were carried out by the author.

%%%%%%%%%%%%%%%%%%%%%%%%%%%%%%%%%%%%%%%%%%%%%%%%%%
\section*{Data Availability}

The numerical data and scripts used to produce the figures in this article will be made available upon reasonable request to the author.

%%%%%%%%%%%%%%%%%%%%%%%%%%%%%%%%%%%%%%%%%%%%%%%%%%
% Appendices
%%%%%%%%%%%%%%%%%%%%%%%%%%%%%%%%%%%%%%%%%%%%%%%%%%
%
% The following appendix placeholders are kept commented for later use.
% Uncomment them only if numerical convergence tests or additional
% parameter-space figures are added before submission.
%
%\appendix
%
%\section{Numerical implementation and convergence tests}
%
%\section{Additional parameter-space figures}

%%%%%%%%%%%%%%%%%%%% REFERENCES %%%%%%%%%%%%%%%%%%

\bibliographystyle{mnras}
\bibliography{1.References}

%%%%%%%%%%%%%%%%%%%%%%%%%%%%%%%%%%%%%%%%%%%%%%%%%%

\bsp
\label{lastpage}
\end{document}